\documentstyle[12pt]{article}
\newcommand{\be}{\begin{equation}}
\newcommand{\ee}{\end{equation}}

\newcommand{\eea}{\end{array}}
\newcommand{\bX}{{\bf X}}
\newcommand{\bN}{{\bf N}}

\title{GENERALIZED WEIERSTRASS-ENNEPER INDUCING,\\
CONFORMAL IMMERSIONS, AND GRAVITY}
\author{Robert Carroll\thanks{Mathematics Department, University
of Illinois, Urbana, IL 61801}\and Boris Konopelchenko\thanks
{Physics Department, University of Lecce, 73100 Lecce, Italy and
Budker Institute of Nuclear Physics, Novosibirsk 90, Russia}}

\date{May, 1995}

\begin{document}

\bibliographystyle{plain}
\maketitle

\begin{abstract}
Basic quantities related to 2-D gravity, such as
Polyakov extrinsic action, Nambu-Goto
action, geometrical action, and Euler characteristic are studied using
generalized Weierstrass-Enneper (GWE) inducing of surfaces in ${\bf R}^3$.
Connection of the GWE inducing
with conformal immersion is made and
various aspects of the theory are shown to be invariant under the modified
Veselov-Novikov hierarchy of flows.  The geometry of $h\sqrt{g} = 1$
surfaces ($h\sim$ mean curvature) is shown to be connected with the
dynamics of infinite and finite dimensional integrable systems.
Connections to Liouville-Beltrami gravity are indicated.
\end{abstract}

\section{INTRODUCTION}

\renewcommand{\theequation}{1.\arabic{equation}}\setcounter{equation}{0}

2-D gravity is one of the most interesting and intriguing toy models
of the last decade.  It has been studied very intensively starting with the
original papers of Polyakov \cite{pe}.  The variety of different approaches
used is rather impressive (see e.g. \cite{gi,hk}).  One of the approaches
consists in the study of Polyakov's surface analogue of the path integral
in terms of original continuous surfaces without discretization,
triangularization, matrix models, etc.  Interesting results in this
direction have been obtained recently in \cite{pa,pc,va,vb} where a theory
of conformal immersion connected with W gravity in the conformal gauge,
strings, and extrinsic geometry has been developed.  In particular, the
importance of 2-D surfaces with $h\sqrt{g} = 1$ was demonstrated where
$h$ is the mean curvature.  An explicit form of the effective action
$\Gamma_{eff}$ for such surfaces was constructed which is a gauge invariant
combination of 2-D intrinsic gravity action in light cone gauge $\Gamma_
{+}$, geometric action a la Virasoro $\Gamma_{-}$, and extrinsic Polyakov
action $\tilde{S}_P$ as in QCD.
\\[3mm]\indent
In the present paper we propose a different approach based on
generalized Weier-\\
strass-Enneper (GWE) inducing.  The method of inducing
surfaces was developed in \cite{kg,kh}.  It allows one to generate
surfaces in ${\bf R}^3$ via simple formulas and to describe their
dynamics via $2+1$ dimensional soliton equations.  The GWE inducing is
a particular case.  In this case the integrable dynamics of surfaces is
generated by the modified Veselov-Novikov (mVN) hierarchy of equations.
We show that GWE inducing is equivalent to the Kenmotsu representation
theorem and establish a correspondence with the conformal immersion
theory.  We express basic quantities of the theory such as Polyakov
extrinsic action, Nambu-Goto action, geometrical action, Euler
characteristic, etc. in terms of basic quantities of the GWE inducing
(two complex variables $\psi_1$ and $\psi_2$).  For compact orientable
surfaces with $h\sqrt{g} = 1$ it is shown that the
Polyakov extrinsic action is invariant under the mVN hierarchy of flows.
We demonstrate that the surfaces with $h\sqrt{g} = 1$ are induced via
the solution of a $1+1$ dimensional Hamiltonian system.  In the one
dimensional limit this system is a dynamical system with four degrees of
freedom which is completely integrable in the Liouville sense.
Connections to Liouville-Beltrami gravity are made in relating
$\Gamma_{eff} = 0$ (corresponding to fixed Euler characteristic $\chi$)
to $\Gamma_{\pm}$ and $\tilde{S}_P$; then the invariance of extremal
$\Gamma_{eff} = 0$ under mVN flows yields a family of extremal
surfaces.

\section{BACKGROUND}
\renewcommand{\theequation}{2.\arabic{equation}}\setcounter{equation}{0}

We will give here some information about the differential geometry of
surfaces, the method of inducing surfaces and their integrable
evolution, and conformal immersion.

\subsection{Surfaces in ${\bf R}^3$}
We consider a surface in the three dimensional Euclidean space
${\bf R}^3$ and will denote the local coordinates of the surface by
$u^1,\,\,u^2$.  The surface can be defined by the formulas (see e.g.
\cite{dm,ed,fo,wi})
\be
\bX^i = x^i(u^1,u^2),\,\,i = 1,2,3
\label{LA}
\ee
where $\bX^i\,\,(i = 1,2,3)$ are the coordinates of the variable point
of the surface and $x^i(u^1,u^2)\,\,(i = 1,2)$ are scalar functions.
The basic characteristics of the surface are given by the first
($\Omega_1$) and second ($\Omega_2$) fundamental forms
\be
\Omega_1 = ds^2 = g_{\alpha\beta}du^{\alpha}du^{\beta};\,\,
\Omega_2 = d_{\alpha\beta}du^{\alpha}du^{\beta}
\label{LB}
\ee
where $g_{\alpha\beta}$ and $d_{\alpha\beta}$ are symmetric tensors and
$\alpha,\,\,\beta$ take the values $1,2$.  Here and below it is assumed
that summation over repeated indices is performed.  The quantities
$g_{\alpha\beta}$ and $d_{\alpha\beta}$ are calculated by the formulas
\be
g_{\alpha\beta} = \frac{\partial\bX^i}{\partial u^{\alpha}}\cdot\frac
{\partial\bX^i}{\partial u^{\beta}};\,\,d_{\alpha\beta} = \frac
{\partial^2\bX^i}{\partial u^{\alpha}\partial u^{\beta}}\cdot\bN^i
\,\,(\alpha,\beta = 1,2)
\label{LC}
\ee
where $\bN^i$ are the components of the normal vector
\be
\bN^i = (det\,g)^{-\frac{1}{2}}\epsilon^{ikm}\frac{\partial\bX^k}
{\partial u^1}\frac{\partial\bX^m}{\partial u^2}\,\,(i = 1,2,3)
\label{LD}
\ee
and $\epsilon^{ikm}$ is a totally antisymmetric tensor with $\epsilon^
{123} = 1$.
The metric $g_{\alpha\beta}$ completely determines the intrinsic
properties of the surface.  The Gaussian curvature $K$ of the surface
is given by the Gauss formula $K = R_{1212}(det\,g)^{-1}$ where
$R_{\alpha\beta\gamma\delta}$ is the Riemann tensor.
\\[3mm]\indent
Extrinsic properties of surfaces are described by the Gaussian curvature
$K$ and the mean curvature $2h = g^{\alpha\beta}d_{\alpha\beta}$.
Embedding of the surface into ${\bf R}^3$ is described both by $g_{\alpha
\beta}$ and $d_{\alpha\beta}$ and it is governed by the Gauss-Codazzi
eequations
\be
\frac{\partial^2\bX^i}{\partial u^{\alpha}\partial u^{\beta}} -
\Gamma^{\gamma}_{\alpha\beta}\frac{\partial\bX^i}{\partial u^{\gamma}}
-d_{\alpha\beta}\bN^i = 0
\label{LE}
\ee
\be
\frac{\partial\bN^i}{\partial u^{\alpha}} + d_{\alpha\gamma}g^{\gamma
\beta}\frac{\partial\bX^i}{\partial u^{\beta}} = 0\,\,
(i = 1,2,3;\,\alpha,\beta = 1,2)
\label{LF}
\ee
where $\Gamma^{\gamma}_{\alpha\beta}$ are the Christofel symbols.
\\[3mm]\indent
Among the global characteristics of surfaces we mention the integral
curvature (see e.g. \cite{dm,ed,fo,wi})
\be
\chi = \frac{1}{2\pi}\int_S K(det\,g)^{\frac{1}{2}}d^2 u
\label{LG}
\ee
where $K$ is the Gaussian curvature and the integration in (\ref{LG})
is performed over the surface.  For compact oriented surfaces
$$\chi = 2(1-g)$$
where $g$ is the genus of the surface and we will generally assume that
surfaces are compact and oriented unless otherwise specified.
\\[3mm]\indent
Families of parametric curves on the surface form a system of curvilinear
local coordinates on the surface.  It is often very convenient to use special
types of parametric curves on surfaces as coordinates.  We will consider
in particular minimal lines (curves of zero length).  In this case
$g_{11} = g_{22} = 0$, i.e.
\be
\Omega_1 = 2g_{12}du^1 du^2
\label{LH}
\ee
For real surfaces minimal lines are complex and $\Omega_1 = 2\lambda
(z,\bar{z})dzd\bar{z}$ where bar means the complex conjugation and
$\lambda$ is a real function.  The Gaussian curvature in this case
is reduced to $K = (1/g_{12})(\partial^2\,log(g_{12})/\partial u^1
\partial u^2)$.

\subsection{Surface evolution}

First we recall the idea of the method of inducing surfaces
following \cite{ki,kg}.
The main idea is to start with a linear PDE $L(\partial_1,
\partial_2)\psi = 0$ in two independent variables $u^1,\,u^2$ with matrix
valued coefficients ($\psi$ is a square matrix).  A formal adjoint
operator $L^{*}$ is obtained via $<\phi,\psi> = \int\int du^1 du^2
Tr(\phi\psi)$ and one has an adjoint equation $L^{*}(\partial_1,
\partial_2)\psi^{*} = 0$.  It follows that
\be
\psi^{*} L\psi - \psi L^{*}\psi^{*} = \partial_1 P_1 - \partial_2 P_2
\label{AA}
\ee
where the $P_i$ are bilinear combinations of $\psi$ and $\psi^{*}$.  Thus
for solutions $\psi,\,\,\psi^{*}$ of $L\psi = 0$ and $L^{*}\psi^{*}
= 0$ one has $\partial_1P_1^{ik} = \partial_2P_2^{ik}$.  This implies
that there exists $w^{ik}$ such that
\be
P_1^{ik} = \partial_2 w^{ik};\,\,P_2^{ik} = \partial_1 w^{ik}
\label{AB}
\ee
and the quantities $X^i = \gamma^{ikj}w^{kj}\,\,(\gamma^{ikj}$ are
constant) given by quadratures
\be
X^i = \gamma^{ikj}\int_{\Gamma}(P_2^{kj} du^1 + P_1^{kj} du^2)
\label{AC}
\ee
do not depend on the curve $\Gamma$.  Now consider quantities of the type
$X^i\,\,(i = 1,2,3)$ as tentative local coordinates of a surface in
${\bf R}^3$ induced by L.  For example any three linearly independent solutions
$\psi_i$ of $L\psi_i = 0$ would induce a tentative surface (for fixed
$\gamma^{ikj}$).  Assume further that the coefficients of L depend on
t and satisfy a t dependent equation
$M(\partial_t,\partial_1,\partial_2)\psi = 0$
for some linear operator M.  Then compatability of the M equation with $L\psi
= 0$ provides a nonlinear PDE for the coefficients of L and we also have
an evolving family of surfaces - provided of course that the coordinate
functions fit together properly to define a surface.
\\[3mm]\indent
The method of inducing surfaces described above is not completely
new.  It is in fact the extension of the ideas of Weierstrass and Enneper
for construction of minimal surfaces (surfaces with $h=0$).  The approach
of Weierstrass and Enneper is as follows.  Let $\phi$ and $\psi$ be
arbitrary functions and define
\be
\partial_z w_1 = i(\psi^2 + \phi^2);\,\,\partial_z w_2 = \psi^2 -
\phi^2;\,\,\partial_z w_3 = -2p\psi;\,\,X^1 = \Re w_1 =
\label{AG}
\ee
$$ \Re\int i(\psi^2 +  \phi^2)dz;\,\,X^2 = \Re w_2 = \Re\int (\psi^2
-\phi^2)dz;\,\,X^3 = \Re w_3 = -\Re\int 2\psi\phi dz$$
Then the $X^i$ define a minimal surface with $z = c$ and $\bar{z} = \hat{c}$
as minimal lines.  Note $\phi$ and $\psi$ are determined via $\partial
_{\bar{z}}\psi = 0,\,\,\partial_{\bar{z}}\phi = 0$.
The straightforward generalization of the Weierstrass-Enneper formulas
to the case of nonminimal surfaces was given in \cite{ki,kg}.  We
start with the system
\be
L\psi =
\left(
\begin{array}{cc}
\partial_z & 0\\
0 & \partial_{\bar{z}}
\end{array}
\right)
\psi +
\left(
\begin{array}{cc}
0 & -p\\
p & 0
\end{array}
\right)
\psi = 0
\label{AH}
\ee
with p real and $\psi$ a $2 \times 2$ matrix.  For $\psi^T =$ transpose
$\psi$ one sees that $\psi^{*}$ satisfies the same equation as $\psi^T$
so $\psi^{*} = \psi^T$ can be stipulated.  Further for $\sigma_2 =
{0 \,\;\; 1 \choose -1 \, 0}$ there is a constraint $\sigma_2\psi
\sigma_2^{-1} = \bar{\psi}$ so a solution of (\ref{AH}) has the form $\tilde
{\psi} = {\psi_1 \, -\bar{\psi}_2 \choose \psi_2 \,\;\; \bar{\psi}_1}$
(we will use $\tilde{\psi}$ now to avoid confusion with $\psi$ of Section 2.3).
For $X^i$ real and $g_{\alpha\beta} = 0\,\,(\alpha \not= \beta)$ one obtains
(cf. \cite{ki,kg})
\be
\partial_z X^1 = i(\psi^2_2 + \bar{\psi}_1^2);\,\,\partial_{\bar{z}}
X^1 = -i(\psi^2_1 + \bar{\psi}^2_2);\,\,\partial_z X^2 = \bar{\psi}_1^2
- \psi^2_2;
\label{AI}
\ee
$$ \partial_{\bar{z}} X^2 = \psi_1^2 - \bar{\psi}^2_2;
\partial_z X^3 = -2\psi_2\bar{\psi}_1;\,\,\partial_{\bar{z}} X^3 = -2\psi_1
\bar{\psi}_2;$$
$$g_{12} = \partial_zX^i\partial_{\bar{z}}X^i =
2(\psi_1\bar{\psi}_1 + \psi_2\bar{\psi}_2)^2 = 2det^2\psi;\,\,
d_{12} = 2p det\psi$$
Further for real $p(z,\bar{z})$ we can write
\be
X^1 + iX^2 = 2i\int_{\Gamma}(\bar{\psi}_1^2dz' - \bar{\psi}_2^2
d\bar{z}');\,\, X^1 -i X^2 =
\label{AJ}
\ee
$$ = 2i\int_{\Gamma}(\psi^2_2dz' - \psi^2_1d\bar{z}');\,\,X^3 =
-2\int_{\Gamma}(\psi_2\bar{\psi}_1dz' + \psi_1\bar{\psi}_2d\bar{z}')$$
where $\Gamma$ is an arbitrary curve in {\bf C} ending at z.
Then $\Omega_1 = 4det^2\tilde{\psi} dzd\bar{z}$ and the
Gaussian and mean curvatures are
\be
K = -det^{-2}\tilde{\psi}(log\,det\tilde{\psi})_{z\bar{z}};\,\,
h = pdet^{-1}\tilde{\psi}
\label{AK}
\ee
and consequently the total curvature is
\be
\chi = \frac{1}{2\pi}\int_{S}\int K\sqrt{det\,g}d^2 u =
\label{AL}
\ee
$$= -\frac{2i}{\pi}\int_{S}\int dz\wedge d\bar{z}(log\,det
\tilde{\psi})_{z\bar{z}} =
\frac{2i}{\pi}\int_{\partial S}dz(log\,det\tilde{\psi})_z$$
(cf. remarks after (\ref{AU}) below).
Hence $\chi$ is determined by the asymptotics of $\psi_1$ and $\psi_2$.
To examine this write (\ref{AH}) as (*) $\psi_{1z} = p\psi_2;\,\,
\psi_{2\bar{z}} = -p\psi_1$ and let $p\to 0$ as $|z|\to \infty$.  Then
$\psi_1\sim a(\bar{z})$ and $\psi_2\sim b(z)$ as $|z|\to\infty$ where
a and b are arbitrary functions.  For solutions of (*) defined by
$|\psi_1|^2\to |z|^n,\,\,\psi_2\to 0$ as $|z|\to\infty$ one obtains
$\chi = -2n$.  Minimal surfaces $\sim p = 0$ and $\psi = \frac{1}
{\sqrt{2}}\psi_2,\,\,\phi = \frac{1}{\sqrt{2}}\bar{\psi}_1$ yields
the Weierstrass-Enneper situation.
As for time evolution with $u^1\sim z,\,\,u^2\sim \bar{z}$ the simplest
nontrivial example is
\be
M(\partial_t,\partial_z,\partial_{\bar{z}}) = \partial_t + \partial_z^3 +
\partial^3_{\bar{z}} + 3{0 \,\, p_z \choose 0 \,\;w}\partial_z
+ 3{\bar{w} \,\;\, 0 \choose p_{\bar{z}} \, \, 0}\partial_{\bar{z}} +
\frac{3}{2}{\bar{w}_{\bar{z}} \,\;\;\; 2pw \choose -2p\bar{w} \,\, w_z}
\label{AM}
\ee
which corresponds to a nonlinear integral equation for p
\be
p_t + p_{zzz} + p_{\bar{z}\bar{z}\bar{z}} + 3p_z w + 3p_{\bar{z}}\bar{w}
+ \frac{3}{2}p\bar{w}_{\bar{z}} + \frac{3}{2}pw_z = 0;\,\, w_{\bar{z}} =
(p^2)_z
\label{AN}
\ee
This equation is the first higher equation in the Davey-Stewartson (DS)
hierarchy for $p,\,q$ with $q = -p$ and it can be connected via a
(degenerate) Miura type transformation with the Veselov-Novikov NVN-II
equation, so one refers to (\ref{AN}) as the modified VN (mVN) equation
(cf. \cite{bj,ki,kg,kj}).  The equations for $\psi_1$ and $\psi_2$ are
given in (\ref{JB}).
\\[3mm]\indent
The hierarchy of integrable PDE associated with the linear problem
(LP) (\ref{AH}) arises as compatibility conditions of (\ref{AH}) with
LP's of the form $\psi_t + A_n\psi = 0;\,\,A_n = \sum_0^n(q_j(u,t)
\partial_{u^1}^{2j+1} + r_j(u,t)\partial_{u^2}^{2j+1})$.  All members
of this mVN hierarchy commute with each other and are integrable by the
inverse scattering method.  Thus the integrable dynamics of surfaces
referred to their
minimal lines is induced by the mVN hierarchy via (\ref{AJ}).  For
such dynamics one is able to solve the initial value problem for the
surface, namely $(g_{\alpha\beta}(z,\bar{z},t=0),d_{\alpha\beta}
(z,\bar{z},t=0))\mapsto (g_{\alpha\beta}(z,\bar{z},t),d_{\alpha\beta}
(z,\bar{z},t))$, using the corresponding results for the equations
from the mVN hierarchy.  This integrable dynamics of surfaces inherits
all properties of the mVN hierarchy.  Note that the minimal surfaces
($p=0$) are invariant under such dynamics.
\\[3mm]\indent For the 1-D limit
one can impose on $p,\,\tilde{\psi}$ the following constraints
$(\partial_z - \partial_{\bar{z}})p = 0;\,\,(\partial_{\bar{z}} -
\partial_z)\tilde{\psi} = 2i\lambda\tilde{\psi}$
($\lambda$ real).  Then $\tilde{\psi}^{*}\,\,(f^{*}\sim\bar{f}$)
satisfies the same constraints and
consequently the $X^k$ are constrained via
$(\partial_{\bar{z}} - \partial_z)X^k = 4i\lambda X^k$
($k = 1,2,3$).  Define now real isometric coordinates $\sigma,\,s$ via
$z = \frac{1}{2}(s - i\sigma)$ to obtain $p = p(s,t),\,\,\tilde{\psi} =
exp(\lambda\sigma)\chi(s,t)$ and $X^k = exp(2\lambda\sigma)\tilde{X}^k
(s,t)\,\,(k = 1,2,3)$.  It follows that $K = 0$ and $K_m = 2p\,exp
(-2\lambda\sigma)$.  These equations describe a cone type surface
generated by the curve with coordinates $\tilde{X}(s,t)$ - i.e. the
surface is effectively reduced to a curve with curvature $p(s,t)$.  The
linear problem (\ref{AH}) is reduced to a 1-D, AKNS type problem for $\chi$
with spectral parameter $\lambda$, i.e.
\be
\partial_s\chi = {i\lambda \,\;\;\; p \choose -p \, -i\lambda}\chi
\label{AO}
\ee
and equation (\ref{AN}) is converted into the mKdV equation $p_t + 2p_{sss}
+ 12p^2p_s = 0$.  Similarly the higher mVN equations pass to higher
order mKdV equations.  In this direction note further
$(\partial_{\bar{z}} - \partial_z)X^k\cdot(\partial_{\bar{z}} -
\partial_z)X^k = -16\lambda^2X^kX^k$.  Via $\partial_zX^k\partial_zX^k =
\partial_{\bar{z}}X^k\partial_{\bar{z}}X^k = 0$ and (\ref{AI}) one obtains
$(2\lambda^2)X^kX^k = det^2\psi$.  But $X^k = exp(2\lambda\sigma)\tilde{X}^k$
and $\psi = exp(\lambda\sigma)\chi$ implies then $(2\lambda^2)\tilde{X}^k
\tilde{X}^k = det^2\chi$.  But for the 1-D
constraint above $det\chi$ = constant (say 1)
which entails then $\tilde{X}^k\tilde{X}^k = (2\lambda)^{-2}$.  Thus
the curve with coordinates $\tilde{X}^k(s,t)$ lies on a sphere of radius
$1/2\lambda$ (as in \cite{do}).  For $\lambda = 0$ one obtains integrable
motions of plane curves as in \cite{gm,gn,lg,nf}.  Note also that (\ref{AC})
implies that tangent vectors to the surface will be expressed in terms
of bilinear combinations of $\psi$ and $\psi^{*}$.

\subsection{Conformal immersions}
We go next to \cite{pa,pc,va,vb} involving conformal immersions and
will sketch some of the results (cf. also \cite{cb,cu,gt,gu,gw,ho,kz,ky,pe,pf,
sg,sh,si,sl,su,sv,ta,wa,ys,zd,ze}).  Consider an oriented 2-D surface
immersed in ${\bf R}^n$, realized as a conformal immersion of a Riemann
surface S, i.e. $X:\,S\to {\bf R}^n$.  This means the induced metric
on S can be written in the form $g_{11} = g_{22};\,\,g_{12} = g_{21} = 0$.
Pick complex local coordinates $z = \xi^1 + i\xi^2$ and $\bar{z} =
\xi^1 - i\xi^2$ so $g_{z\bar{z}} = g_{\bar{z}z} \not= 0$ and $g_{zz} =
g_{\bar{z}\bar{z}} = 0$.  The Grassmannian $G_{2,n}$ of oriented 2-planes
in ${\bf R}^n$ can be represented by the complex quadric $Q_{n-2}$ in
${\bf C}P^{n-1}$ defined by $\sum_1^n w^2_k = 0,\,\,w_k\in {\bf C}$
where $w_k\,\,( k = 1,...,n)$ are homogeneous coordinates in ${\bf C}P^{n-1}$.
Writing $w_k = a_k + ib_k$ with $\vec{A} = \{a_k\},\,\,\vec{B} =
\{b_k\},\,\,Q_{n-2}$ involves $\| \vec{A}\| = \|
\vec{B}\|$ with $\vec{A}\cdot\vec{B} = 0$.  Then $\vec{A},\,\vec{B}$
form the basis for an oriented 2-plane in ${\bf R}^n$ and an $SO(2)$
rotation of vectors gives rise to the same point $\{exp(i\theta)w_k\}$ in
${\bf C}P^{n-1}$.  In the conformal gauge above the tangent plane to S
spanned by $(\partial_{\xi^1}X^{\mu},\partial_{\xi^2}X^{\mu})$
corresponds to the point $(\partial_{\xi^1}X^{\mu} + i\partial_{\xi^2}
X^{\mu})\sim \partial_{\bar{z}}X^{\mu}\in Q_{n-2}$.  The (conjugate)
Gauss may is defined now by
$\bar{G}(z) = [\partial_z X]$.
Thus $S\to G_{2,n}\simeq Q_{n-2}$ and one looks for a function $\psi(z,
\bar{z})$ (to be determined) such that $\partial_z X^{\mu} = \psi\phi^{\mu}$
where $\phi^{\mu}\in Q_{n-2}$ satisfies $\phi^{\mu}\phi_{\mu} = 0$
($\phi:\,\,S\to Q_{n-2}$).
Note that a map $S\to Q_{n-2}$ corresponds locally to $\phi(z,\bar{z}) =
(\phi_1,...,\phi_n)\in {\bf C}^n/0$ satisfying $\sum_1^n\phi^2_k = 0$
(here $S\to G_{2,n}\sim (\xi^1,\xi^2)
\simeq (z,\bar{z})\to (\partial_{\xi^1} X^{\mu},\partial_{\xi^2} X^{\mu})$
while $G_{2,n}\simeq Q_{n-2}$ via ($\partial_{\xi^1} X^{\mu},\partial_
{\xi^2}X^{\mu})\simeq (\partial_z X^{\mu})$).  The line element in S
is $ds^2 = \lambda^2|dz|^2$ where $\lambda^2 = 2\|\partial_z X
\|^2 = 2|\psi|^2\|\phi\|^2\,\,(\|\phi\|^2 = \phi^{\mu}\bar{\phi}_{\mu})$
and the mean curvature
vector field of S is $H^{\mu} = (2/\lambda^2)X^{\mu}_{z\bar{z}}$ (note
$\phi^{\mu}$ is tangent to S and $H^{\mu}$ is normal).  To see this
one uses the Gauss-Codazzi equations in the form (cf. (\ref{LE}))
\be
\partial_{\alpha}\partial_{\beta}X = \Gamma^{\gamma}_{\alpha\beta}
\partial_{\gamma}X + H^i_{\alpha\beta}N_i;\,
\partial_{\alpha}N_i = -H^i_{\alpha\beta}g^{\beta\gamma}\partial_
{\gamma} X + (N_j\cdot\partial_{\alpha}N_i)N_j
\label{AP}
\ee
where $\Gamma^{\gamma}_{\alpha\beta}\sim$ affine connection determined
by the induced metric $g_{\alpha\beta}$ and $H^i_{\alpha\beta}\,\,
(i = 1,...,n-2)$ are the components of the second fundamental form along
the $n-2$ independent normals $N_i$.  Then one notes that in the conformal
gauge only $\Gamma^z_{zz}$ and $\Gamma^{\bar{z}}_{\bar{z}\bar{z}}$ are
nonzero and $H = \frac{1}{2}H^{i\alpha}_{\alpha}N_i$.  Assuming $\psi$
exists it can be determined as follows (cf. also \cite{ho}).  Express
$H^{\mu}$ in terms of $\psi$ and $\phi^{\mu}$ and write
\be
(log\psi)_{\bar{z}} = -\eta;\,\,V^{\mu}\equiv\partial_{\bar{z}}\phi^{\mu}
-\eta\phi^{\mu} = \bar{\psi}\|\phi\|^2H^{\mu};\,
\eta = (\partial_{\bar{z}}\phi^{\mu})\bar{\phi}^{\mu}/\|\phi\|^2
\label{AQ}
\ee
Here $\|\phi\|^2 = \phi^{\mu}\bar{\phi}_{\mu}$ and $V^{\mu}$ is the normal
component of $\partial_{\bar{z}}\phi^{\mu}$.  Since $H^{\mu}$ and $\|\phi\|^2$
are real, $V^{\mu}$ can be written $V^{\mu} = exp(i\alpha)R^{\mu}$ for
$R^{\mu}$ real with $\alpha$ the argument of $\bar{\psi}$ for $\psi =
\rho exp(-i\alpha)$.  The first two equations in (\ref{AQ}) are the
integrability conditions on the Gauss map (not every $G_{2,n}$ field
$\phi^{\mu}$ forms a tangent plane to a given
surface).  Now $V^{\mu}$ is a linear
combination of $n-2$ unit normals to S and $V^{\mu} = exp(i\alpha)R^{\mu}$
so there are n-3 conditions here plus a remaining condition determined
by $\Re\eta_z = -(log\rho)_{z\bar{z}}$ and $\alpha_{z\bar{z}} =
\Im\eta_z$.
\\[3mm]\indent Now we concentrate on ${\bf R}^3$ although many results
for ${\bf R}^n$ appear in \cite{pa} for example (cf. also \cite{kz}).
Thus $G_{2,3}\simeq
Q_1\simeq{\bf C}P^1\simeq S^2$ and the Gauss map can be expressed by a
single complex (Kaehler) function $f(z,\bar{z})$ via
\be
\phi = (1-f^2,i(1+f^2),2f);\,\,\phi^{\mu}\phi_{\mu} = 0;
\label{AR}
\ee
$$ \|\phi\|^2 = 2(1 + |f|^2)^2;\,\,or\,\,via\,\,N = \frac{1}
{1 + |f|^2}(f+\bar{f},-i(f - \bar{f}),|f|^2 -1)$$
($N\sim$ normal Gauss map) and the integrability condition $\Im\eta_z =
\alpha_{z\bar{z}}$ is given by
\be
\Im[(f_{z\bar{z}}/f_{\bar{z}}) - (2\bar{f}f_z/(1 + |f|^2))] = 0
\label{AS}
\ee
One obtains then
\be
V^{\mu} = -2f_{\bar{z}}N^{\mu};\,\,h = H^{\mu}N_{\mu}\,\,or\,\,
H^{\mu} = hN^{\mu};\,\,\psi = -\bar{f}_z/h(1 + |f|^2)^2
\label{AT}
\ee
(h is the mean curvature scalar).  It follows then that
$(log\psi)_{\bar{z}} = -2\bar{f}f_{\bar{z}}/(1 + |f|^2)$.
{}From this one computes the Euler characteristic ($\chi(g) = 2(1-g)$)
\be
2\pi\chi(g) = \int\sqrt{g}Rd^2\xi = 2\int\frac{|f_{\bar{z}}|^2 -
|f_z|^2}{(1 + |f|^2)^2}\,\frac{i}{2}dz\wedge d\bar{z}
\label{AU}
\ee
Note here that (\ref{AU}) is expressed via globally defined objects
whereas (\ref{AL}) requires e.g. $det\tilde{\psi}\not= 0$ or $\infty$.
We will see that for $h\sqrt{g} = 1$ surfaces $det\tilde{\psi} = (1/2p)$
so, assuming $p\not=\infty$ at interior points and that $p$ has bounded
derivatives $p_z,\,\,p_{\bar{z}},\,\,p_{z\bar{z}}$ in the interior, one
can only use (\ref{AL}) when $p\not= 0$ at interior points.  Since this
could preclude some interesting situations
we will use (\ref{AU}) for calculation and refer to this as $\chi$
throughout.
The Polyakov action (or action induced by external curvature) is
\be
\tilde{S}_P = \frac{2}{g_0^2}\int \|H\|^2\sqrt{g}\,d^2\xi = \frac{2}{g_0^2}
\int\frac{\|V\|^2}{\|\phi\|^2}\frac{i}{2}dz\wedge d\bar{z}
= \frac{4}{g_0^2}\int\frac{|f_{\bar{z}}|^2}{(1 + |f|^2)^2}\frac{i}
{2}dz\wedge d\bar{z}
\label{AV}
\ee
and the Nambu-Goto action is ($S_{NG} = \sigma\int\sqrt{g}d^2\xi$)
\be
S_{NG} = \sigma\int|\psi|^2\|\phi\|^2
\frac{i}{2}dz\wedge d\bar{z}
= 2\sigma\int\frac{|f_{\bar{z}}|^2}{h^2(z,\bar{z})(1 + |f|^2)^2}
\frac{i}{2}dz\wedge d\bar{z}
\label{AW}
\ee
We will be concerned mainly with $\tilde{S}_P$.
\\[3mm]\indent A special role is played by surfaces where $h\sqrt{g} = c$.
Thus we will introduce a local orthonormal moving frame $\hat{e}_1,
\,\,\hat{e}_2,$ and $\hat{N}\sim \hat{e}_3$ where
$\hat{e}_1,\hat{e}_2$ are tangent
to S and $\hat{N}$ is normal.  One can choose e.g.
\be
\hat{e}_1 = \frac{1}{1+|f|^2}(1-\frac{1}{2}(f^2+\bar{f}^2),\frac{i}{2}
(f^2 - \bar{f}^2),f+\bar{f});\,\,\hat{e}_2 =
\label{AX}
\ee
$$ \frac{1}{1+|f|^2}(\frac{i}{2}(f^2 - \bar{f}^2),1 + \frac{1}{2}(f^2 +
\bar{f}^2),-i(f-\bar{f}));\,\,\hat{N} = \frac{1}{1 + |f|^2}(f+\bar{f},
-i(f - \bar{f}), |f|^2 - 1)$$
The structural equations (\ref{AP}) take the form
\be
\partial_z
\left(
\begin{array}{c}
\hat{e}_1\\
\hat{e}_2\\
\hat{e}_3
\end{array}
\right) = A_z
\left(
\begin{array}{c}
\hat{e}_1\\
\hat{e}_2\\
\hat{e}_3
\end{array}
\right);
\label{AY}
\ee
%$$ A_z =
%\left(
%\begin{array}{ccc}
%0 & \frac{-i(f\bar{f}_z - \bar{f}f_z)}{1+|f|^2} & \frac{-(f_z-\bar{f}_z)}
%{1 + |f|^2}\\
%\frac{i(f\bar{f}_z - \bar{f}f_z)}{1+|f|^2}
%& 0 & \frac{i(f_z - \bar{f}_z)}{1+|f|^2}\\
%\frac{f_z - \bar{f}_z}{1+|f|^2} & \frac{-i(f_z-\bar{f}_z)}{1+|f|^2} & 0
%\end{array}
%\right)$$
where $A_z$ is a matrix involving $f,\,\bar{f},\,f_z,\,\bar{f}_z$.
There will also be an analogous identical equation involving $\partial/
\partial\bar{z}$.  Thus $\partial_z\hat{e}_i = (A_z)_{ij}\hat{e}_j,\,\,
\partial_{\bar{z}}\hat{e}_i = (A_{\bar{z}})_{ij}\hat{e}_j$.
Then $A_z,\,A_{\bar{z}}$ are components of a vector
$\vec{A}$ in conformal gauge which transforms as a 2-D $SO(3,{\bf C})$
gauge field.  Under a local gauge transformation $\hat{e}^T\to
g\hat{e}^T,\,\,S\to S',$ and $A\to A'$ where in an obvious
notation (dropping the arrows) $A'_{\pm} = gA_{\pm}g^{-1} -(\partial_{\pm}g)
g^{-1}\,\,(g\in SO(3,{\bf C})$).  Using the $SO(2)$ degree of freedom
involved in choosing the $\hat{e}_i$ to rotate away a component $A_{12}$
of the tangential connection via
$g_0(\psi)$ one arrives at
\be
A'_z =
\left(
\begin{array}{ccc}
0 & 0 & \frac{1}{\sqrt{2}}(\frac{H_{zz}}{\sqrt{g}} + h\sqrt{g})\\
0 & 0 & \frac{i}{\sqrt{2}}(\frac{-H_{zz}}{\sqrt{g}} + h\sqrt{g})\\
-\frac{1}{\sqrt{2}}(\frac{H_{zz}}{\sqrt{g}}+h\sqrt{g}) &
-\frac{i}{\sqrt{2}}(-\frac{H_{zz}}{\sqrt{g}}+h\sqrt{g}) & 0
\end{array}
\right);
\label{AZ}
\ee
$$ A'_{\bar{z}} =
\left(
\begin{array}{ccc}
0 & -i[(log\bar{\psi})_{\bar{z}} + \frac{2f\bar{f}_{\bar{z}}}{1+|f|^2}] &
\frac{1}{\sqrt{2}}(H_{\bar{z}\bar{z}} + h)\\
i[(log\bar{\psi})_{\bar{z}} + \frac{2f\bar{f}_{\bar{z}}}{1+|f|^2}] & 0 &
\frac{i}{\sqrt{2}}(H_{\bar{z}\bar{z}} - h)\\
-\frac{1}{\sqrt{2}}(H_{\bar{z}\bar{z}}+h) &
-\frac{i}{\sqrt{2}}(H_{\bar{z}\bar{z}}
-h) & 0
\end{array}
\right)$$
Here one has used $H_{zz} = H^{\mu}_{zz}N^{\mu} = -2f_z\psi$ where
$h = -\bar{f}_z/\psi(1+|f|^2)^2$ along with $\sqrt{g} = |\psi|^2
\|\phi\|^2 = 2|\psi|^2(1+|f|^2)^2$.  This transformation resembles
\cite{su,sv} but works at a deeper level since $\psi$ is involved (cf.
\cite{pc}).  Further argument with currents and a gauge fixing leaves
$T_{zz}$ unfixed and
the condition $h\sqrt{g} = 1$
(or any constant) then singles out a certain class of surfaces
(cf. also \cite{su,sv} where light cone gauge is used).  In
the conformal gauge $\sqrt{g} = exp(\xi)$ where $\xi$ is the Liouville
mode.  In particular the Polyakov action $\tilde{S}_P$, or extrinsic
geometrical
action (\ref{AV}), can be considered
as a gauge fixed form (in conformal gauge) of the action
\be
\hat{S}_P = \frac{i}{g_0^2}\int\sqrt{h}h^{\alpha\beta}
\frac{\partial_{\alpha}f\partial_{\beta}\bar{f}}{(1+|f|^2)^2}\,
dz\wedge d\bar{z}
\label{BA}
\ee
(this is the same as (\ref{AV}) plus terms modulo Euler characteristic
as will be shown later).  The
EM tensor for (\ref{BA}) is $T_{zz} = -\partial_z
f\partial_z\bar{f}/(1+|f|^2)^2$
so using $H_{zz} = -2f_z\psi, \,\,h = -\bar{f}_z/\psi(1+|f|^2)^2,\,\,
\sqrt{g} = 2|\psi|^2(1+|f|^2)^2$, we get $T_{zz} = (H_{zz}/\sqrt{g})
(h\sqrt{g}) = H_{zz}/\sqrt{g}$ when $h\sqrt{g} = 1$.
For constant $h\sqrt{g},\,\,(A'_{\bar{z}})_{12}$ in (\ref{AZ}) becomes
$i\partial_{\bar{z}}log(h) = -i\xi_{\bar{z}}\,\,(\sqrt{g} = exp(\xi))$
yielding the transformation for the induced metric
in Polyakov's 2-D gravity so $H_{\bar{z}\bar{z}}\sim$ induced metric
for surfaces of constant $h\sqrt{g}$ while $T_{zz} = H_{zz}/\sqrt{g}\sim$
EM tensor.
Finally one notes that $h\sqrt{g} =$ constant surfaces
are characterized by $\psi f_{\bar{z}} =$ constant (cf. (\ref{MH})).
\\[3mm]\indent  Now  \cite{va}, which begins with a summary of
the ${\bf R}^3$ situation just discussed, provides further information.
Thus first in summary, if one uses $H_{\bar{z}\bar{z}}$ and $H_{zz}/
\sqrt{g}$ as independent dynamical degress of freedom (independent of the
$X^{\mu}$ variables) then the integrability condition $\partial_{\bar{z}}
A'_z - \partial_zA'_{\bar{z}} + [A'_z,A'_{\bar{z}}] = 0$ can be rewritten
with $\xi$ or directly as an equation of motion
\be
\partial^3_zH_{\bar{z}\bar{z}} = (\partial_{\bar{z}} - H_{\bar{z}\bar{z}}
\partial_z - 2(\partial_zH_{\bar{z}\bar{z}}))(H_{zz}/\sqrt{g})
\label{BB}
\ee
Some useful formulas involving $H_{\bar{z}\bar{z}}$ and $H_{zz}/\sqrt{g}$
for $h\sqrt{g} = 1$ can now be obtained as follows.
Thus from $\partial_zX^{\mu} = \psi\phi^{\mu},
\,\,\phi^{\mu} = (1-f^2,i(1+f^2),2f),\,\,\psi = -\partial_z\bar{f}/
h(1+|f|^2)^2,$ and $h\sqrt{g} = 1$ one finds $(\dagger)\,\,\psi
\partial_{\bar{z}}f = -\frac{1}{2}$ while from the Gauss-Codazzi equations
(\ref{AP}) $H_{\bar{z}\bar{z}} = -2\bar{\psi}\partial_{\bar{z}}\bar{f}$.
This plus $(\dagger)$ yields
\be
H_{\bar{z}\bar{z}} = \partial_{\bar{z}}\bar{f}/\partial_z\bar{f}
\label{BC}
\ee
Further for $h\sqrt{g} = 1$ the integrability condition (\ref{AS}) can be
simplified via $(\dagger)$ and $(log\psi)_{\bar{z}} = -\eta$ to
\be
f_{\bar{z}\bar{z}}/f_{\bar{z}} - 2\bar{f}f_{\bar{z}}/(1+|f|^2) = 0
\label{BD}
\ee
This implies
\be
T_{zz} = H_{zz}/\sqrt{g} = \frac{\partial^3_z\bar{f}}{\partial_z\bar{f}}
-\frac{3}{2}\big(\frac{\partial^2_z\bar{f}}{\partial_z\bar{f}}\big)^2
= D_z\bar{f}
\label{BE}
\ee
Note here that in (\ref{BC}) $H_{\bar{z}\bar{z}}$ has the form of a Beltrami
coefficient $\mu = \bar{\partial}\bar{f}/\partial\bar{f}$ and
$T_{zz}$ is the corresponding Schwartz derivative.  Thus an equation
(\ref{BB}) becomes $\partial^3\mu = \bar{\partial}T_{zz} - \mu\partial T_{zz} -
2(\partial\mu)T_{zz}$.
Now one notes also that the independent dynamical fields
$H_{\bar{z}\bar{z}}$ and $H_{zz}/\sqrt{g}$ can be parametrized in terms of
independent Gaussian maps as $H_{\bar{z}\bar{z}} = \partial_{\bar{z}}
f_2/\partial_z f_2$ and $H_{zz}/\sqrt{g} = D_z f_1$ (the $f_i$ determine
the image of the $X^{\mu}$ in $G_{2,3}$).  Then in in \cite{va}
an effective action depending on $f_1,\,\,f_2$ is determined and the
equation of motion (\ref{BB}) is used to constrain these fields.
First one derives an action invariant under Virasoro symmetries (since
$h\sqrt{g} = 1$ surfaces have Virasoro symmetry following earlier
remarks - cf. here also \cite{pe}).  The gauge invariant action
$\Gamma_{eff}$ depends on $A_z$ and $A_{\bar{z}}$ (we omit the $'$ now)
via parametrizations $A_z = u^{-1}\partial_z u$ and $A_{\bar{z}} =
v^{-1}\partial_{\bar{z}}v$.  Here $u,v$ are independent elements of the
gauge group and this will correspond to taking $H_{\bar{z}\bar{z}}$ and
$H_{zz}/\sqrt{g}$ as independent of $X^{\mu}$.  Now write (cf. \cite{pe})
\be
\Gamma_{eff} = \Gamma_{-}(A_z) + \Gamma_{+}(A_{\bar{z}}) -
\frac{k}{4\pi}Tr\int A_z A_{\bar{z}} dz\wedge d\bar{z}
\label{BF}
\ee
where $k = n_f =$ the number of fermions and
\be
\Gamma_{-}(A_z) = \frac{k}{8\pi}Tr\int[(\partial_z u)u^{-1}(\partial_
{\bar{z}} u)u^{-1}]d^2\xi +
\label{BG}
\ee
$$ +\frac{k}{12\pi}Tr\int\epsilon^{abc}
(\partial_a u)u^{-1}(\partial_b u)u^{-1}(\partial_c u)u^{-1}d^2\xi dt$$
Then $\Gamma_{+}(A_{\bar{z}})$ is given by a similar expression with
$u\to v$ and the sign changed in the last integral (cf. also \cite{pe}.
Now take $A_z^{+} = H_{zz}/\sqrt{g},\,\,A_z^{-} \equiv h\sqrt{g} = 1,$ and
$A^0_z = 0$ to get (cf. (\ref{AZ}) - factors of $i/2$ are being dropped
in integration)
\be
\Gamma_{-}(A_z) = S_{-}(f_1) = \frac{k}{8\pi}\int\frac{\partial_{\bar{z}}
f_1}{\partial_z f_1}[\frac{\partial^3_z f_1}{\partial_z f_1} -
2(\frac{\partial^2_z f_1}{\partial_z f_1})^2] dz\wedge d\bar{z}
\label{BH}
\ee
This corresponds to geometrical action (cf. \cite{ag,ap,aq,cb,cu}).
Calculation of $\Gamma_{+}(A_{\bar{z}})$ from (\ref{AZ}) is not so easy but
in light cone gauge an explicit determination is possible, leading to
\be
S_{+}(f_2) = -\frac{k}{8\pi}\int\frac{\partial^2_z f_2}{\partial_z f_2}
(\frac{\partial_z\partial_{\bar{z}}f_2}{\partial_z f_2} -
\frac{\partial^2_z f_2\partial_{\bar{z}}f_2}{\partial_z f_2\partial_z f_2})
dz\wedge d\bar{z}
\label{BI}
\ee
This is exactly the form of the light cone action in 2-D intrinsic gravity
theory.  Finally the total action on
$h\sqrt{g} = 1$ surfaces is
\be
\Gamma_{eff}(f_1,f_2) = \frac{k}{8\pi}\int\frac{\partial_{\bar{z}} f_1}
{\partial_z f_1}[\frac{\partial^3_z f_1}{\partial_z f_1} -
2(\frac{\partial^2_z f_1}{\partial_z f_1})^2]
dz\wedge d\bar{z}\,\, -
\label{BJ}
\ee
$$ -\frac{k}{8\pi}\int\frac{\partial^2_z f_2}{\partial_z f_2}
(\frac{\partial_z\partial_{\bar{z}} f_2}{\partial_z f_2} -
\frac{\partial^2_z f_2\partial_{\bar{z}} f_2}{(\partial_z f_2)^2})
dz\wedge d\bar{z} -\frac{k}{4\pi}\int\frac{\partial_{\bar{z}}f_2}
{\partial_z f_2}D_z f_1 dz\wedge d\bar{z}$$
This is the extrinsic geometric gravitational WZNW action on $h\sqrt{g} = 1$
surfaces in light cone gauge.  It combines in a gauge
invariant way the geometric and light cone action studied in 2-D intrinsic
gravity.
\\[3mm]\indent The equation of motion for (\ref{BJ}) is
\be
\partial^3_z(\frac{\partial_{\bar{z}} f_2}{\partial_z f_2})
- \partial_{\bar{z}}D_z f_1 - (\frac{\partial_
{\bar{z}} f_2}{\partial_z f_2})\partial_zD_z f_1 -
2\partial_z(\frac{\partial_{\bar{z}} f_2}{\partial_z f_2})
D_z f_1 = 0
\label{BK}
\ee
obtained by varying $f_1$ and $f_2$ independently, and one can see that
this is equivalent to (\ref{BB}) which can be regarded as relating
$H_{\bar{z}\bar{z}}$ and $H_{zz}/\sqrt{g}$.  It is automatically
satisfied when one takes both $T_{zz}$ and $H_{\bar{z}\bar{z}}$ as
determined by extrinsic geometry via $X^{\mu}$.
Now one wants an effective
action in terms of $H_{\bar{z}\bar{z}}$ and $H_{zz}/\sqrt{g}$ through
their parametrizations in terms of the $f_i$ such that these fields are
independent of $X^{\mu}$.  First one checks that (\ref{BJ}) is invariant
under Virasoro transformations.  Next one shows that
$\Gamma_{eff}(f_1,f_2) = \Gamma_{eff}(f_1\circ f,f_2\circ f)$ and chooses
$f = f_2^{-1}(z,\bar{z})$ where $f_2(f_2^{-1}(z,\bar{z}),\bar{z}) = z$
so $\Gamma_{eff}(f_1,f_2) = \Gamma_{-}(f_1\circ f_2^{-1}) = \Gamma_{+}
(f_2\circ f_1^{-1})$ (the last by interchanging $f_1,f_2$).  This leads to
\be
\Gamma_{+}(f_2\circ f_1^{-1}) = \Gamma_{-}(f_1\circ f_2^{-1}) =
\Gamma_{+}(f_2) + \Gamma_{-}(f_1) - \frac{k}{4\pi}\int\frac{\partial_{\bar{z}}
f_2}{\partial_z f_2}D_z f_1 dz\wedge d\bar{z}
\label{BL}
\ee
Thus in particular the properties of $\Gamma_{eff}$ can be understood from
either $\Gamma_{+}(f_2\circ f_1^{-1})\sim$ light cone action of intrinsic
2-D gravity (with $f\sim f_2\circ f_1^{-1}$) or from $\Gamma_{-}(f_1\circ
f_2^{-1})\sim$ geometric action arising from quantization of the Virasoro group
by coadjoint orbits.  The last (coupling) term corresponds exactly to
the extrinsic Polyakov action $\tilde{S}_P$ modulo $\chi$ (cf. Theorem 4.5).
In fact the coupling term in $\Gamma_{eff}$ is needed in
order to make it invariant under Virasoro transformations of $F_1,\,
F_2$ (recall $H_{\bar{z}\bar{z}} = \mu(F_2)$ and $D_z F_1 = T_{zz} =
H_{zz}/\sqrt{g}$).  Quantization in the $\bar{z}$ sector is developed
after modification of the conformal weight of $F_1\circ F_2^{-1}$
(where one is thinking of the geometric action realization).  $\Gamma_{eff}$
is the conformally invariant extension of $\tilde{S}_P$ where $T_{zz}$
and $H_{\bar{z}\bar{z}}$ are the dynamical fields.  There is also a hidden
Virasoro symmetry on $h\sqrt{g} = 1$ surfaces where $H_{\bar{z}\bar{z}}$
and $T_{zz}$ transform as a metric and an energy momentum tensor respectively
under Virasoro action.  The Gauss map is important in establishing the
existence
of the Virasoro symmetry in $h\sqrt{g} = 1$ surfaces.

\subsection{Comments on geometry and gravity}

We make here a few further comments about the Liouville
equation, Liouville action, etc.  The Liouville equation classically
involves e.g. $\partial^2_{z\bar{z}}\phi = -\frac{1}{2}Kexp(2\phi)$
or (for $\rho = exp(2\phi)\,\,\partial^2_{z\bar{z}} log(\rho) = -K\rho$
where $K\sim$ Gaussian curvature.  On the other hand classical conformal
unquantized Liouville action involves e.g. ($\gamma\sim\hbar$)
\be
S_L = \frac{1}{4\pi\gamma^2}\int\sqrt{\hat{g}}(\frac{1}{2}\hat{g}^
{ab}\partial_a\phi\partial_b\phi + \phi R(\hat{g}) + \frac{\mu}{2}e^{\phi})
\label{EG}
\ee
$$ =\frac{1}{4\pi}\int\sqrt{\hat{g}}(\frac{1}{2}\hat{g}^{ab}\partial_a
\phi\partial_b\phi + \frac{1}{\gamma}\phi R(\hat{g}) + \frac{\mu}
{2\gamma^2}e^{\gamma\phi})$$
as in \cite{gi} (cf. also \cite{cu,hk} for other notations).  Note that
the second formula follows from the first via $\phi\to\gamma\phi$.  The
equations of motion from (\ref{EG}) are evidently
\be
\gamma\Delta\phi = R(\hat{g}) + \frac{\mu}{2}e^{\gamma\phi}
\label{EH}
\ee
and from \cite{gi} $R(exp(2\sigma)\hat{g}) = exp(-2\sigma)[R(\hat{g}) -
2\Delta\sigma)]$.  Hence for $\hat{g}\to g = exp(2\sigma)\hat{g}$ and
$2\sigma = \gamma\phi$ one has $0 = R(g) + (\mu/2)$ or $R(g) =
R(exp(\gamma\phi)\hat{g}) = -(\mu/2)$.  Thus the Liouville field
$\phi$ or $\gamma\phi$ is thrown into the metric and one looks for
a metric with constant Ricci curvature $-(\mu/2)$.  Thus Liouville theory
can be thought of as a theory of metrics and and equation such as
(\ref{EH}) is sometimes called a Liouville equation.
\\[3mm]\indent
Now we know that the Liouville equation with $g = \hat{g}
exp(\gamma\phi)$ provides constant curvature $R_g = -\mu/2$ (given a
background metric $\hat{g}$).  One has equations of motion of the form
($\gamma = 1$) $\xi_{z\bar{z}}\sim\Delta\xi = R(\hat{g}) + (\mu/2)
exp^{\xi}$ as above.
However we must not confuse this with the siutation of
\cite{pa,pc,va,vb} where one should emphasize in particular that the
Polyakov action of (\ref{AV}) or (\ref{BA}) is a special action
introduced for QCD to cope with quantum fluctuations.  It becomes
the kinetic energy term of a Grassmannian sigma model (cf. \cite{vb,za}
where the Nambu-Goto action or area term also becomes an action with
local coupling $1/h^2$).  It is not the same as the Polyakov action of
Liouville gravity, which is equivalent to the Nambu-Goto action there,
but rather a string theoretic term in QCD (as well as a crucial geometric
ingredient for W gravity).  This is related to the idea that a geometric
realization of W gravity as extended 2-D gravity involves, in ${\bf R}^3$,
surfaces of constant mean curvature density ($h\sqrt{g} = 1$) in which
$T_{zz}\sim (H_{zz}/\sqrt{g})$.  The corresponding W algebra in this case
is the Virasoro algebra.  This is accomplished in a conformal gauge
for the induced metric ($\sim H_{\bar{z}\bar{z}}$).  The mathematics
however, involving the Kaehler function f of (\ref{AR}), then leads to
formulas similar to those of Liouville-Beltrami intrinsic gravity a la
\cite{cu,gt,gu,gw,pe,pf,sg,sh,si,sl,wa} for example (e.g. formulas such as
(\ref{BB}), (\ref{BH}), (\ref{BI}), etc.).  In particular the Polyakov action
$\tilde{S}_P$ or $\hat{S}_P$
leads to the basic EM tensor $T_{zz}$ and metric $H_{\bar{z}\bar{z}}$
above which can be used as basic variables (via Kaehler functions f)
in formulating an effective action $\Gamma_{eff}$ as in (\ref{BJ}).
\\[3mm]\indent Further, following
\cite{pc}, one has to be careful to distinguish conformal gauge and
light cone gauge  (cf. here \cite{su,sv} where
light cone gauge is used).  Also we must recall that in \cite{pc},
the condition $h\sqrt{g} = 1$ is a gauge fixing, and some formulas hold
more generally before such a fixing.  For example the Gauss-Codazzi
equations imply $H_{\bar{z}\bar{z}} = -2\bar{\psi}\partial_{\bar{z}}
\bar{f}$ and in general one has also (cf. equations after (\ref{BA}) -
this is organized in Section 3).
\be
\sqrt{g} = 2|\psi|^2(1 + |f|^2)^2;\,\,H_{zz} = -2f_z\psi;\,\,h =
-\frac{\bar{f}_z}{\psi(1 + |f|^2)^2}
\label{FB}
\ee
On the other hand after gauge fixing, $h\sqrt{g} = 1$, one has
(cf. Section 3)
\be
\psi\partial_z f = -\frac{1}{2};\,\,T_{zz} = H_{zz}/\sqrt{g}
\label{FC}
\ee
The formula $T_{zz} = H_{zz}/\sqrt{g}$ arises after gauge fixing but is
not itself a fixing (cf. \cite{pc,va}).
One notes also that $h\sqrt{g} = 1$ is the conformal
analogue of the condition $h = 1$ of \cite{su,sv} where light cone gauge
is used with $\sqrt{g} = 1/4$ (in conformal gauge $\sqrt{g} = exp(\xi)$
where $\xi$ is the Liouville mode or field).
Similarly in \cite{su,sv} one uses $T_{zz}\sim H_{zz}$.  The role of
$H_{\bar{z}\bar{z}}$ as induced metric corresponds then
(for $h\sqrt{g} = 1$) to $\mu
= \bar{\partial}\bar{f}/\partial\bar{f}$ being the
induced metric.  Equations such as (\ref{BB}) take the form then
\be
\partial^3\mu = [\bar{\partial} - \mu\partial - 2(\partial
\mu)]T_{zz}\;\;(T_{zz} = \frac{H_{zz}}{\sqrt{g}})
\label{FD}
\ee
and as in (\ref{BE}) for $h\sqrt{g} = 1$ we have $T_{zz} = D_z\bar{f}$.
Such formulas also arise in \cite{gt,gu,gw,sg,sh,si,sl,wa} (cf. \cite{cu})
and we will look at this below.  We will want to compare now various formulas
for various actions involving Beltrami coefficients (divergence terms are
frequently added without changing the theory).

\section{CONNECTING GWE INDUCING AND\newline CONFORMAL IMMERSIONS}
\renewcommand{\theequation}{3.\arabic{equation}}\setcounter{equation}{0}

We refer here also to \cite{cu} where some preliminary calculations were
made.

\subsection{Relations between quantities}

It is clear that there is a strong interaction between
the material just sketched on induced surfaces
and conformal immersions; we will establish
some precise connections here.  This will provide some new
relations between integrable systems and gravity theory.  First
consider (cf. (\ref{AI}))
\be
\partial_z X^{\mu} = (i(\psi^2_2 + \bar{\psi}^2_1),\bar{\psi}_1^2 -
\psi^2_2,-2\psi_2\bar{\psi}_1)
\label{DA}
\ee
Evidently $\partial_z X^{\mu}\cdot\partial_z X^{\mu} = 0$ with
\be
\|\partial_z X^{\mu}\|^2 = \partial_z X^{\mu}\cdot\bar{\partial_z X^{\mu}}
= 2(|\psi_1|^2 + |\psi_2|^2)^2 = 2det^2\tilde{\psi}
\label{DB}
\ee
($\tilde{\psi}$ will be used for the matrix involving $\psi_1,\,\,\psi_2$
and $\psi$ will be used in $\partial_z X^{\mu} = \psi\phi^{\mu}$).  We
note that the Weierstrass-Enneper (WE) ideas motivated work of Kenmotsu
\cite{ky} which underlies some work on the Gauss map (cf. \cite{ho})
so there are natural background connections here (some of this
is spelled out later).  Now let $\phi^{\mu}$
coordinatize the map $S\to Q_1$ and be represented by (\ref{AR}) for some
complex function f.  We can determine f and $\psi$ in terms of $\psi_1,\psi_2$
via
\be
i(\psi^2_2 + \bar{\psi}_1^2)= \psi(1 - f^2);\,\,\bar{\psi}_1^2 -
\psi^2_2 = i\psi(1 + f^2);\,\,-2\psi_2\bar{\psi}_1 = 2\psi f
\label{DC}
\ee
This gives
\\[3mm]\indent{\bf THEOREM 3.1}.$\,\,$ GWE inducing (\ref{AH}), (\ref{AJ})
and the Kenmotsu representation are equivalent and one has
\be
f = i\bar{\psi}_1/\psi_2;\,\,\psi = i\psi_2^2
\label{DD}
\ee
\indent {\it Proof}.$\,\,$
{}From the formulas
\be
(log\psi)_{\bar{z}} = -\frac{2\bar{f}f_{\bar{z}}}{1 + |f|^2};\,\,h = -\frac
{\bar{f}_z}{\psi(1+|f|^2)^2}
\label{MA}
\ee
for real $h$ one gets
$\bar{\psi}\bar{f}_z = \psi f_{\bar{z}};\,\,\psi_{\bar{z}}f = \bar{\psi}_z
\bar{f}$.  The Kenmotsu theorem gives the condition
\be
h[f_{z\bar{z}} - \frac{2\bar{f}f_zf_{\bar{z}}}{1+|f|^2}] = h_zf_{\bar{z}}
\label{QQ}
\ee
for existence of a surfaces with a given Gauss map and mean curvature $h$
and (\ref{MA}) with its conclusion correspond to this (cf. \cite{ho},
second reference).
Writing now
\be
\psi_1 = -\bar{f}\sqrt{-i\bar{\psi}};\,\,\psi_2 = \sqrt{-i\psi};\,\,
p = -\frac{\psi f_{\bar{z}}}{|\psi|(1+|f|^2)}
\label{MD}
\ee
one shows that equations (\ref{MA}) and its conclusion are equivalent to the
system $L\psi = 0$ of (\ref{AH}), or $\psi_{1z} = p\psi_2;\,\,
\psi_{2\bar{z}} = -p\psi_1$.  Evidently (\ref{DD}) holds and the
Jacobian of the transformation $(f,\psi)\to (\bar{\psi}_1,\psi_2)$ is
equal to 2.  {\bf QED}
\\[3mm]\indent
We will now develop some relations between the $\psi_i,\,\,p,\,\,\psi,$
and $f$.  Situations arising from the constraint $h\sqrt{g} = 1$ will
be distinguished from the general case when possible,
 but the derivations are often
run together for brevity.  The situations of most interest here involve
$h\sqrt{g} = 1$ and we will therefore concentrate on this.
First in general, from (\ref{AK}),
one has $h = pdet^{-1}\tilde{\psi}$ and
$ds^2 = \lambda^2 dzd\bar{z}$ where $\lambda^2 = 2\|\partial_z X^{\mu}\|^2$
(we choose this definition for $\lambda$ and will change
symbols for other $\lambda$).  Hence from (\ref{DB})
\be
\lambda^2 = 4det^2
\tilde{\psi},\,\,det\tilde{\psi} = \lambda/2;\,\,h =
2p/\lambda
\label{MG}
\ee
Note also the agreement of K in \cite{ki,pa}.
Now recall (after (\ref{BA})), $h = -\bar{f}_z/\psi(1 + |f|^2)^2$ and $\sqrt{g}
=2|\psi|^2(1 + |f|^2)^2$ so $h\sqrt{g} = -2\bar{f}_z|\psi|^2/\psi =
-2\bar{f}_z\bar{\psi}$ and since $\overline{(f_z)} = \bar{f}_{\bar{z}}$,
one has
\be
h\sqrt{g}=1\equiv \psi f_{\bar{z}} = -1/2
\label{MH}
\ee
Also for $h\sqrt{g} = 1$ from
(\ref{BE}) $T_{zz} = H_{zz}/\sqrt{g} = D_z\bar{f}$ and the integrability
condition (\ref{AS}) takes the form (\ref{BD}).  This leads to ($\psi_{1z} =
p\psi_2\sim \bar{\psi}_{1\bar{z}} = p\bar{\psi}_2,\,\,\psi_{2\bar{z}} =
-p\psi_1\sim \bar{\psi}_{2z} = -p\bar{\psi}_1$)
\be
\psi f_{\bar{z}} = -\frac{1}{2} = i\psi^2_2(i\bar{\psi}_1/\psi_2)_{\bar{z}}
= -\psi^2_2(\frac{\bar{\psi}_{1\bar{z}}}{\psi_2} - \frac{\bar{\psi}_1
\psi_{2\bar{z}}}{\psi_2^2})
\label{DE}
\ee
$$ = -p(|\psi_1|^2 + |\psi_2|^2)\Rightarrow det\tilde{\psi} = |\psi_1|^2
+ |\psi_2|^2 = 1/2p\,\,\,(h\sqrt{g} = 1)$$
Putting this in $det\tilde{\psi} = \lambda/2$ gives $\lambda = 1/p$ and
$h= 2p^2 = 2/\lambda^2$ while $K = -det^{-2}\tilde{\psi}
(logdet\tilde{\psi})_{z\bar{z}} = -4p^2(logdet\tilde{\psi})_{z\bar{z}}$
(note $h\sqrt{g} = c$ is of interest here - not $h = c$).
We also write ($\bar{f} = -i\psi_1/\bar{\psi}_2,\,\,
\partial_{\bar{z}}\bar{f} = -(i/\bar{\psi}_2^2)(\bar{\psi}_2\psi_{1\bar{z}}
- \psi_1\bar{\psi}_{2\bar{z}})$)
\be
H_{\bar{z}\bar{z}} = \frac{\bar{f}_{\bar{z}}}{\bar{f}_z} =
\frac{\bar{\psi}_2\psi_{1\bar{z}} - \psi_1\bar{\psi}_{2\bar{z}}}
{p(|\psi_1|^2 + |\psi_2|^2)} = 2(\bar{\psi}_2\psi_{1\bar{z}}
-\psi_1\bar{\psi}_{2\bar{z}})= 2\bar{\psi}_2^2\partial_{\bar{z}}
(\frac{\psi_1}{\bar{\psi}_2})\,\,(h\sqrt{g} = 1)
\label{DF}
\ee
Also from (\ref{BE}), noting that $\partial_z\bar{f} = -ip(|\psi_1|^2
+ |\psi_2|^2)/\bar{\psi}_2^2 = -i/2\bar{\psi}_2^2$, which implies
$\partial_z^2\bar{f} = (i/2)2\bar{\psi}_{2z}/\bar{\psi}_2^3 =
-ip\bar{\psi}_1/\bar{\psi}_2^3$ and $\partial_z^3\bar{f} = -ip_z\bar{\psi}_1/
\bar{\psi}_2^3 - ip\bar{\psi}_{1z}/\bar{\psi}^3_2 - 3ip^2\bar{\psi}_1^2/
\bar{\psi}_2^4$, one obtains
\be
T_{zz} = \frac{2}{\bar{\psi}_2^2}(p_z\bar{\psi}_1\bar{\psi}_2 +
p\bar{\psi}_2\bar{\psi}_{1z}) = \frac{2}{\bar{\psi}_2}\partial_z
(p\bar{\psi}_1)\,\,\,(h\sqrt{g} = 1)
\label{DG}
\ee
\\[3mm]\indent {\bf PROPOSITION 3.2}.$\,\,$ For $h\sqrt{g} = 1$ we have
(\ref{DE}) - (\ref{DG}).
\\[3mm]\indent {\bf REMARK 3.3}.$\,\,$
We indicate here some calculations designed in particular to confirm
various results in \cite{pc}.  Thus for $h\sqrt{g} = 1$ we have first
(recall $\overline{(F_z)} = \bar{F}_{\bar{z}}$)
\be
f = \frac{i\bar{\psi}_1}{\psi_2};\,\psi = i\psi^2_2;\,\psi_{1z} =
p\psi_2;\,\bar{\psi}_{1\bar{z}} = p\bar{\psi}_2;\,\psi_{2\bar{z}} =
-p\psi_1;\,\bar{\psi}_{2z} = -p\bar{\psi}_1;
\label{FA}
\ee
$$ \sqrt{g} = e^{\xi} = |\psi|^2\|\phi\|^2 = 2|\psi|^2(1+|f|^2)^2 =
2det^2\tilde{\psi} = \frac{\lambda^2}{2};\,\lambda = \frac{1}{p};\,
h = 2p^2$$
Recall next that $h\sqrt{g} = 1\sim \psi f_{\bar{z}} = -1/2$ and from
$\sqrt{g} = 2|\psi|^2(1+|f|^2)^2$ one gets $h = 2/\lambda^2 = 2p^2$
(also $h = -\bar{f}_{\bar{z}}/[\psi(1+|f|^2)^2]$ - cf. (\ref{MA})).  From
$H_{zz} = -2f_z\psi = -(2f_z/f_{\bar{z}})f_{\bar{z}}\psi$ and (\ref{DF}),
namely $H_{\bar{z}\bar{z}} = \bar{f}_{\bar{z}}/\bar{f}_z$, we see that
$H_{\bar{z}\bar{z}} = \overline{(H_{zz})}$.  Note that in general one
expects only $\overline{(H_{zz})} = \overline{((H_z)_z)} =
\overline{(H_z)}_{\bar{z}} = \bar{H}_{\bar{z}\bar{z}}$.  Further from \cite{ki}
$K = -4p^2(log det\tilde{\psi})_{z\bar{z}} = -4p^2(log(1/2p))_{z\bar{z}}\,\,
(1/2p = |\psi_1|^2 + |\psi_2|^2$).  Evidently $(log h)_{\bar{z}} =
-\xi_{\bar{z}}$.  Further $\sqrt{g} = exp(\xi) = 1/2p^2$ implies $2p^2 =
exp(-\xi) = h$ and $\xi = -log(2p^2)$ with $\xi_{z\bar{z}} = -2(log p)_
{z\bar{z}}$ while $K = -4p^2 [log(1/2p)]_{z\bar{z}}$ implies $K =
2exp(-\xi)(log p)_{z\bar{z}}$ so $ K = 2exp(-\xi)(-\xi_{z\bar{z}}/2) =
-\xi_{z\bar{z}} exp(-\xi)$ and hence $\xi_{z\bar{z}} = -Kexp(\xi)$
or $\xi_{z\bar{z}} = -K/h = -K\sqrt{g}$.
Note that in \cite{pc} one writes $\xi_{z\bar{z}} = Kexp(-\xi)$ which is
equivalent to $(-\xi)_{z\bar{z}} = -Kexp(-\xi)$ or $\xi_{z\bar{z}} =
-Kexp(\xi)$.  Also we have for
$h\sqrt{g} = 1$ the equations
$\psi\bar{\partial}f = -(1/2)$ and this with $h = -(\bar{f}_z/\psi
(1 + |f|^2)^2)$ implies $h = (2\partial\bar{f}\bar{\partial}f/(1 + |f|^2)^2)$
while in general $H_{\bar{z}\bar{z}} = -2\bar{\psi}\bar{f}_{\bar{z}}$
plus $h\sqrt{g} = 1$ implies $H_{\bar{z}\bar{z}} = \bar{\partial}\bar{f}/
\partial\bar{f}$ (cf. (\ref{BC}).
\\[3mm]\indent
We want to exhibit next the restrictions (if any)
on $p,\,\psi_1,\,\psi_2$ which
are imposed by the requirements (2.28), $h\sqrt{g} = 1$ and $T_{zz}
= H_{zz}/\sqrt{g}$ (note (\ref{DG}) is the calculation $T_{zz} = D_z\bar{f}$
and $H_{zz} = -2\psi f_z = -2i\psi^2_2\partial(i\bar{\psi}_1/
\psi_2) = 2(\bar{\psi}_{1z}\psi_2 - \bar{\psi}_1\psi_{2z})$).
One obtains first then $T_{zz} = H_{zz}/\sqrt{g} = 4p^2(\bar{\psi}_{1z}\psi_2
- \bar{\psi}_1\psi_{2z})$ which must equal $(2/\bar{\psi}_2)
\partial(p\bar{\psi}_1)$ by (\ref{DG}).  Hence we have the following
conditions on $p,\,\psi_1,\,\psi_2$
\be
2p^2(\bar{\psi}_{1z}\psi_2 - \bar{\psi}_1\psi_{2z}) = \frac{1}{\bar{\psi}_2}
\partial(p\bar{\psi}_1);
\label{FF}
\ee
$$ |\psi_1|^2 + |\psi_2|^2 = \frac{1}{2p};\,\,\psi_{1z} = p\psi_2;\,\,
\psi_{2\bar{z}} = -p\psi_1$$
(the latter equations being equivalent to $\bar{\psi}_{1\bar{z}} =
p\bar{\psi}_2$ and $\bar{\psi}_{2z} = -p\bar{\psi}_1$) plus
(\ref{BB}) (which will turn out not to be a restriction).
Recall also
\be
\mu = H_{\bar{z}\bar{z}} = \frac{\bar{\partial}\bar{f}}{\partial\bar{f}}
= -2\bar{\psi}\bar{f}_{\bar{z}} =
2(\bar{\psi}_2\psi_{1\bar{z}} - \psi_1\bar{\psi}_{2\bar{z}})
\label{FG}
\ee
which leads to $T_{zz} = 2p^2\bar{\mu}$ which is quite pleasant and
equation (\ref{BB}) has the form $\partial^3\mu =
[\bar{\partial} - \mu\partial - (2\partial\mu)]T_{zz}$.
One can now show with a little calculation that (\ref{FF}) and (\ref{FG})
are compatible and we have the result
\\[3mm]\indent {\bf THEOREM 3.4}.$\,\,$ Given the basic evolving surface
equations $\psi_{1z} = p\psi_2$ and $\psi_{2\bar{z}} = -p\psi_1$ with
p real one achieves a fit with comformal immersions
via (\ref{DD}).  The condition $h\sqrt{g} = 1$ implies then that
$det\tilde{\psi} = |\psi_1|^2 + |\psi_2|^2 = (1/2p)$ (and $h = 2p^2$)
and these conditions imply the first equation of (\ref{FF}) which says
that $H_{zz}/\sqrt{g} = T_{zz} = D_z\bar{f}$.  This all implies $T_{zz} = 2p^2
\bar{\mu}$ ($T_{zz} = H_{zz}/\sqrt{g},\,\,\mu = \bar{f}_{\bar{z}}/
\bar{f}_z$) and the only additional condition then on $p,\,\psi_1,\,
\psi_2$ is that (\ref{BB}) hold in the form $\partial^3\mu =
[\bar{\partial} - \mu\partial - (2\partial\mu)](2p^2\bar{\mu})$.
However this equation
is always true when $T_{zz} = D_z\bar{f}$ with $\mu = \bar{f}_
{\bar{z}}/\bar{f}_z$ a Beltrami coefficient (as is the case here).  This
is stated e.g. in \cite{gt,wa} and verified in \cite{cu}
and below in Section 4 (it is also
implicit in \cite{cb,da}).  This means that (\ref{BB}) is automatically
true and hence there are no a priori
restrictions on $\psi_i,\,\,p$ imposed by
the fit above, beyond the condition $det\tilde{\psi} = 1/2p$.
The Liouville equation $\xi_{z\bar{z}} = -Kexp(\xi)$ also holds automatically
here as do the equations (cf. \cite{cu,pc}) $\partial\mu +
\bar{\partial}\xi = 0$ and $\bar{\partial}\bar{\mu} + \partial\xi = 0$.
\\[3mm]\indent {\it Proof}:$\,\,$ All that remains are the last two
equations which arise in \cite{pc} when $\sqrt{g} = exp(\xi)$ and the
second fundamental form $(H_{\alpha\beta})$ are used as independent
variables.  We check these as follows.  Since $2p^2 = exp(-\xi)$ one has
$-\xi = log(2) + 2log(p)$ so the requirement involves $2log(p)_z =
\bar{\partial}\mu$ and $2log(p)_{\bar{z}} = \partial\mu$.  Then from
$\mu = 2(\bar{\psi}_2\psi_{1\bar{z}} - \psi_1\bar{\psi}_{2\bar{z}})$
we get for example
\be
\mu_z = 2(\bar{\psi}_{2z}\psi_{1\bar{z}} + \bar{\psi}_2\psi_{1z\bar{z}}
-\psi_{1z}\bar{\psi}_{2\bar{z}} - \psi_1\bar{\psi}_{2z\bar{z}}
\label{FK}
\ee
$$ = 2[-p\bar{\psi}_1\psi_{1\bar{z}} + \bar{\psi}_2(p_{\bar{z}}\psi_2 +
p\psi_{2\bar{z}}) -p\psi_2\bar{\psi}_{2\bar{z}} + \psi_1(p_{\bar{z}}
\bar{\psi}_1 + p\bar{\psi}_{1\bar{z}})]$$
$$= 2\{p_{\bar{z}}(|\psi_2|^2 + |\psi_1|^2) - p\bar{\psi}_1\psi_{1\bar{z}}
-p\psi_2\bar{\psi}_{2\bar{z}} + p[\bar{\psi}_2(-p\psi_1) + \psi_1(p
\bar{\psi}_2)]\}$$
Now the last $[\;\;]$ is zero and from (\ref{FF}) we have
$\bar{\psi}_1\psi_{1\bar{z}} + \bar{\psi}_{2\bar{z}}\psi_2 =
-(p_{\bar{z}}/2p^2)$ which implies $\mu_z = 2log(p)_{\bar{z}}$.  The
equation $\bar{\mu}_{\bar{z}} = 2log(p)_z$ is
then automatic. {\bf QED}

\subsection{Expressions and behavior for the actions}

We consider next the various actions in terms of the
$\psi_i$.  Thus from (\ref{AV}) ($f = i\bar{\psi}_1/\psi_2,\,\, f_{\bar{z}} =
ip(|\psi_1|^2 + |\psi_2|^2)/\psi_2^2,\,\,h\sqrt{g} = 1$)
\be
\tilde{S}_P = \frac{2i}{g_0^2}\int\frac{|f_{\bar{z}}|^2}{(1+|f|^2)^2}
dz\wedge d\bar{z} = \frac{2i}{g_0^2}\int p^2 dz\wedge d\bar{z} =
\frac{i}{2g_0^2}\int \frac{dz\wedge d\bar{z}}{(|\psi_1|^2 +
|\psi_2|^2)^2}
\label{DH}
\ee
while from (\ref{BH}) the geometrical action with $f_1 = \bar{f}$ becomes
(cf. calculations in (\ref{DG}))
\be
S_{-} = \frac{k}{8\pi}\int\frac{\partial_{\bar{z}}\bar{f}}
{\partial_z\bar{f}}[\frac{\partial_z^3\bar{f}}{\partial_z\bar{f}} -
2(\frac{\partial_z^2\bar{f}}{\partial_z\bar{f}})^2]dz\wedge d\bar{z} =
\label{DI}
\ee
$$ = \frac{k}{4\pi}\int [(\partial_z(p\bar{\psi}_1)/\bar{\psi}_2) -
p^2(\bar{\psi}_1/\bar{\psi}_2)^2]dz\wedge d\bar{z}$$
(one notes that calculation with $\bar{f}$ is appropriate since
$\mu,\,T$ are defined via $f_1,\,f_2\sim\bar{f}$).
{}From (\ref{BH})
and (\ref{DI}) we can write now
$$
\frac{\partial_z(p\bar{\psi}_1)}{\bar{\psi}_2} - p^2(\frac{\bar{\psi}_1}
{\bar{\psi}_2})^2 = -\frac{\bar{\psi}_{2zz}}{\bar{\psi}_2} -(\frac
{\bar{\psi}_{2z}}{\bar{\psi}_2})^2$$
\be
= -\partial^2log\bar{\psi}_2
-2(\partial log\bar{\psi}_2)^2
\label{GI}
\ee
This leads to
\be
S_{-} = -\frac{k}{4\pi}\int [\partial^2 log\bar{\psi}_2 + 2(\partial
log\bar{\psi}_2)^2]dz\wedge d\bar{z}
\label{GJ}
\ee
We consider also the Nambu-Goto action of (\ref{AW}), which we
write as
$$S_{NG} = \sigma\int\sqrt{g}d^2\xi
= \frac{i\sigma}{2}\int |\psi|^2\|\phi\|^2 dz\wedge d\bar{z} =
i\sigma\int |\psi|^2(1 + |f|^2)^2 dz\wedge d\bar{z}$$
\be
= i\sigma\int (|\psi_1|^2 + |\psi_2|^2)^2 dz\wedge d\bar{z} =
\frac{i\sigma}{4}\int \frac{dz\wedge d\bar{z}}{p^2}
\label{GK}
\ee
Further in general
we look at
\be
S_{+} = -\frac{k}{8\pi}\int\frac{\bar{f}_{zz}}{\bar{f}_z^2}[\bar{f}_
{z\bar{z}} - \frac{\bar{f}_{zz}\bar{f}_{\bar{z}}}{\bar{f}_z}]dz\wedge d\bar{z}
\label{GV}
\ee
and recall however that $\mu = (\bar{f}_{\bar{z}}/\bar{f}_z)$ so
$\mu_z = (\bar{f}_{z\bar{z}}/\bar{f}_z) - (\bar{f}_{\bar{z}}\bar{f}_{zz}/
\bar{f}^2_z)$ while $\mu_z = 2(log(p))_{\bar{z}}$ as well.  Also
\be
\frac{\bar{f}_{zz}}{\bar{f}_z} = \frac{-ip\bar{\psi_1}/\bar{\psi}_2^3}
{-i/2\bar{\psi}_2^2}=-2\frac{\bar{\psi}_{2\bar{z}}}{\bar{\psi}_2} =
-2\bar{\partial}log\bar{\psi}_2
\label{GW}
\ee
Consequently one has
\be
S_{+} = \frac{k}{4\pi}\int (log\bar{\psi}_2)_{\bar{z}}(log(p))_{\bar{z}}
\label{GX}
\ee
Finally we compute also $\chi(g)$ via (\ref{AU}) to get
\be
2\pi\chi(g) = \int R\sqrt{g}d^2\xi = i\int p^2[1 -
|\mu|^2]dz\wedge d\bar{z}
\label{DJ}
\ee
Thus we can state
\\[3mm]\indent {\bf THEOREM 3.5}$\,\,$ For $h\sqrt{g} = 1$
the quantities $\tilde{S}_P,\,\,S_{-},\,\,S_{NG},\,\,S_{+},$
and $\chi$ are given via
(\ref{DH}), (\ref{GJ}), (\ref{GK}), (\ref{GX}), and  (\ref{DJ}).
\\[3mm]\indent
We remark in passing that the genus of our immersed surface corresponds
to the degree of the mapping $S\to {\bf C}P^1$
and the total curvature is $\chi = 2-2g$
For immersions into
${\bf R}^3$ this is the only topological invariant whereas for
${\bf R}^4$ one obtains the Whitney self-intersection number, which
has an interpretation in QCD (cf. \cite{vb}).  See here also the
discussion in \cite{kj} (second book), pp. 169 and 181, in connection
with charge and the Ishimori equation, and Remark 5.4.
\\[3mm]\indent
Consider now the extrinsic Polyakov and Nambu-Goto actions
(cf. (\ref{AV}) - (\ref{DH}) and (\ref{AW}) -
(\ref{GK})) which we rewrite here as ($h\sqrt{g} = 1$)
\be
\tilde{S}_P = \frac{2i}{g_0^2}\int p^2 dz\wedge d\bar{z};\,\, S_{NG} =
\frac{i\sigma}{4}\int\frac{dz\wedge d\bar{z}}{p^2}
\label{JA}
\ee
Now go to the modified Veselov-Novikov (mVN) equations based on (\ref{AM})
to obtain for $M\psi = 0$
\be
\psi_{1t} + \psi_{1zzz} + \psi_{1\bar{z}\bar{z}\bar{z}} + 3\bar{w}\psi_
{1\bar{z}} + \frac{3}{2}\bar{w}_{\bar{z}}\psi_1 + 3(\frac{\psi_{1z}}
{\psi_2})_z\psi_{2z} + 3w\psi_{1z} = 0
\label{JB}
\ee
$$\psi_{2z} + \psi_{2zzz} + \psi_{2\bar{z}\bar{z}\bar{z}} + 3w\psi_{2z} +
3\bar{w}\psi_{2\bar{z}} + \frac{3}{2}w_z\psi_2 + 3(\frac{\psi_{1z}}
{\psi_2})_{\bar{z}}\psi_{1\bar{z}} = 0$$
where $w_{\bar{z}} = -[(\psi_{1z}\psi_{2\bar{z}})/(\psi_1\psi_2)]_z$.  From
the mVN equation (\ref{AN}) one has also
\be
(p^2)_t + (2p_{zz} - p^2_z + 3p^2w)_z + (2pp_{\bar{z}\bar{z}} - p^2_{\bar{z}}
+ 3p^2\bar{w})_{\bar{z}} = 0
\label{JC}
\ee
Consequently we obtain (assume a closed surface or zero boundary terms)
\be
\frac{d\tilde{S}_P}{dt} = \frac{2i}{g_0^2}\int\frac{\partial(p^2)}
{\partial t}dz\wedge d\bar{z} = 0
\label{JD}
\ee
Thus $\tilde{S}_P$ is invariant under the mVN deformations which means
there is an infinite family of suraces with the same $\tilde{S}_P$.  In
particular this would apply to minimal $\tilde{S}_P$ surfaces which in
the corresponding quantum problem would correspond to zero modes.  Further
one knows that the integrals of motion are common for the whole mVN
hierarchy (where the $n^{th}$ time variable would correspond e.g. to
$M_n\sim\partial_t + \partial_z^{2n+1} + \partial_{\bar{z}}^{2n+1} +
\cdots$).  In the one dimensional limit this hierarchy is reduced to the
mKdV hierarchy.  In any event we can state
\\[3mm]\indent {\bf THEOREM 3.6}.$\,\,$ For compact oriented surfaces
$\tilde{S}_P$ is invariant under the whole
mVN hierarchy of deformations ($h\sqrt{g} = 1$).
\\[3mm]\indent  We note however that separately
(\ref{JC}) does not yield zero for $\partial_t S_{NG}$ or $\partial_t
S_{-}$.
\\[3mm]\indent
{}From the point of view of inducing surfaces one continues to
ask what is the role
of the condition $h\sqrt{g} = 1$ and this has the following features.
Thus consider $\psi_{1z} = p\psi_2;\,\,\psi_{2\bar{z}} = -p\psi_1$
under the constraint $|\psi_1|^2 + |\psi_2|^2 = 1/2p$, which leads to
\be
\psi_{1z} -\frac{1}{2}(\frac{\psi_2}{|\psi_1|^2 + |\psi_2|^2}) = 0;\,\,
\psi_{2\bar{z}} + \frac{1}{2}(\frac{\psi_1}{|\psi_1|^2 + |\psi_2|^2}) = 0
\label{JE}
\ee
This system has several simple properties.  First it is Lagrangian wih
Lagrangian
\be
{\cal L} = \psi_1\bar{\psi}_{2z} + \bar{\psi}_1\psi_{2\bar{z}} -
\psi_2\bar{\psi}_{1\bar{z}} - \bar{\psi}_2\psi_{1z} + log(|\psi_1|^2
+ |\psi_2|^2)
\label{JF}
\ee
(confirmation is immediate).
Introducing coordinates $z = (x+ iy)/2$ one has the system
\be
\psi_{1x} - i\psi_{1y} -\frac{1}{2}(\frac{\psi_2}{|\psi_1|^2 + |\psi_2|^2})
= 0;\,\,
\psi_{2x} + i\psi_{2y} + \frac{1}{2}(\frac{\psi_1}{|\psi_1|^2 + |\psi_2|^2})=0
\label{JG}
\ee
where $x$ plays the role of time.  This system has 4 real integrals of
motion, namely
\be
C_{+} = \int dy(\psi_1^2 + \psi_2^2 + \bar{\psi}_1^2 + \bar{\psi}_2^2);\,\,
C_{-} = i\int dy(\psi_1^2 + \psi_2^2 -\bar{\psi}_1^2 - \bar{\psi}_2^2);
\label{JH}
\ee
$$P = \int dy(\psi_{1y}\bar{\psi}_2 - \bar{\psi}_1\psi_{2y});\,\,
{\cal H} = \int dy\{i(\psi_{1y}\bar{\psi}_2 + \psi_{2y}\bar{\psi}_1)
+\frac{1}{2}log(|\psi_1|^2 + |\psi_2|^2)\}$$
Again confirmation is straightforward (note $P_x = -(1/2)\int dy
\partial_y log(|\psi_1|^2 +|\psi_2|^2)$ and ${\cal H}_x = \int dy\cdot 0$).
Next we see that the system can be represented in the Hamiltonian form
\be
\psi_{1x} = \{\psi_1,{\cal H}\};\,\,\psi_{2x} = \{\psi_2,{\cal H}\}
\label{JI}
\ee
where the Poisson brackets are given via
\be
\{f,g\} = \int dy[\frac{\delta f}{\delta\psi_1}\frac{\delta g}{\delta
\bar{\psi}_2} - \frac{\delta f}{\delta\psi_2}\frac{\delta g}{\delta
\bar{\psi}_1} -(f\leftrightarrow g)]
\label{JJ}
\ee
The corresponding symplectic form is $\Omega = d\psi_1\wedge d\bar{\psi}_2
+ d\bar{\psi}_1\wedge d\psi_2$.  The equations (\ref{JI}) are easily
checked and we omit calculations.  One can also say that the interaction
part of the Hamiltonian ${\cal H}$ is
\be
{\cal H}_{int} = \frac{1}{2}log\,det\,\tilde{\psi}
\label{JK}
\ee
which has a pleasant appearance.  Thus (cf. Theorem 3.4)
\\[3mm]\indent {\bf THEOREM 3.7}.$\,\,$ For $h\sqrt{g} = 1$ we have
(\ref{JE}) - (\ref{JK}).
Thus (\ref{JE}) is a Hamiltonian-Lagrangian system
inducing surfaces with the property
$h\sqrt{g} = 1$ via (\ref{AJ}).
\\[3mm]\indent
Let us next consider particular classes of surfaces with $h\sqrt{g} = 1$
which are generated by the Weierstrass-Enneper formulas in the case
$p_z = p_{\bar{z}}$ (one dimensional limit corresponding to curves).
In this case ($z = (1/2)(x+iy)$) referring to \cite{kg} we can write
\be
\psi_1 = r(x)e^{\lambda y};\,\,\psi_2 = s(x)e^{\lambda y}
\label{JL}
\ee
where $r,s$ are complex valued functions and $\lambda = i\mu$ is imaginary.
The system (\ref{JG}) becomes now
\be
r_x + \mu r -\frac{1}{2}(\frac{s}{|r|^2 + |s|^2}) = 0;\,\,
s_x - \mu s + \frac{1}{2}(\frac{r}{|r|^2 + |s|^2}) = 0
\label{JM}
\ee
We write $r = r_1 + ir_2, \,\,s = s_1 + is_2$ then to obtain
($\Xi = r_1^2 + r_2^2 + s_1^2 + s_2^2$)
\be
r_{1x} + \mu r_1 -\frac{s_1}{2\Xi} = 0;\,\,
r_{2x} + \mu r_2 - \frac{s_2}{2\Xi} = 0;
\label{JN}
\ee
$$s_{1x} - \mu s_1 + \frac{r_1}{2\Xi} = 0;\,\,
s_{2x} - \mu s_2 + \frac{r_2}{2\Xi} = 0$$
It is easily checked that this system has the following two integrals
of motion
\be
{\cal H} = -\mu(r_1s_1 + r_2s_2) + \frac{1}{4}log(\Xi);\,\,
M = r_1s_2 - r_2s_1
\label{JO}
\ee
Further the system (\ref{JN}) is Hamiltonian with
\be
r_{ix} = \{r_i,{\cal H}\};\,\,s_{ix} = \{s_i,{\cal H}\}
\label{JP}
\ee
where the Poisson brackets arise from (\ref{JJ}) in the form
\be
\{f,g\} = \int dy[\frac{\delta f}{\delta r_1}\frac{\delta g}{\delta s_1}
- \frac{\delta f}{\delta s_2}\frac{\delta g}{\delta r_2}
- (f\leftrightarrow g)]
\label{JQ}
\ee
One checks that ${\cal H}$ and $M$ are in involution ($\{{\cal H},M\}
=0$) and thus the system (\ref{JN}) is integrable in the Liouville sense
with two degrees of freedom.  The induced Weierstrass-Enneper surfaces
(developable surfaces generated by the curves)
then have the form
$$
X^1 + iX^2 = 2ie^{-2i\mu y}\int[(r_1-ir_2)^2 - (s_1 - is_2)^2]dx';$$
\be
X^3 = -2\int(r_1s_1 + r_2s_2)dx' - 2My
\label{JR}
\ee
and we refer to \cite{kl} for more on this.  In particular we have
(cf. the end of Section 2.2)
\\[3mm]\indent {\bf THEOREM 3.8}.$\,\,$ For $h\sqrt{g} = 1$ and
$p_z = p_{\bar{z}}$ with $\psi_{\bar{z}} - \psi_z = 2i\lambda\psi,\,\,
\lambda$ real, we obtain (\ref{JL}) - (\ref{JR}).

\section{LIOUVILLE-BELTRAMI GRAVITY}
\renewcommand{\theequation}{4.\arabic{equation}}\setcounter{equation}{0}

We recall also some results from \cite{gt,wa}
(we write here $T$ for $T_{zz}$ at times).
The presentation in
\cite{gt,wa} is somewhat abbreviated and unclear at times and we give
here an enhanced treatment of this material with proofs in order to
utilize some of the results later and also to make propaganda for these
matters.  We are led to consider the subject as follows.  One
always will have (\ref{BB})
or (\ref{FC}) when $\mu$ is a Beltrami coefficient and $T$ is the
corresponding Schwartzian.  If $\mu = \delta H/\delta T$ for some
Hamiltonian $H$ then the equation (\ref{FC}) for example becomes a
Hamiltonian equation $\bar{\partial}T = \{T,H\}$.  Such an $H$ can be
constructed in light cone gauge in the form of geometric action
(cf. Remark 4.2 below).
Now it is
asserted in \cite{gt,va} that the coupling term in (\ref{BJ})
corresponds exactly to the extrinsic Polyakov action $\tilde{S}_P$
and we note that this term has the form
\be
S_{int} = -\frac{k}{4\pi}\int \mu(\bar{f})D_z\bar{f}dz\wedge d\bar{z}
\label{GY}
\ee
when $f_1 = f_2 = \bar{f}$ (see below for proof).
Further following \cite{gt,wa} the
Polyakov light cone intrinsic action arises by a simple calculation
from such an $H$ via use of $\int \mu T_{zz}$ (see Remark 4.2 below).  Thus
if ${\it ga}\sim$ geometric action
density and ${\it ipa}\sim$ intrinsic Polyakov action, then the
connection is ${\it ipa} = \mu T - {\it ga}$ (cf.
below for details).
This says that in (\ref{BL}) for $f_1 = f_2 = \bar{f}$
the last two terms represent minus the intrinsic Polykov action and
the extrinsic Polyakov action is simply the sum of the
geometric action and the intrinsic Polyakov action.  Actually the
corresondence here is precise modulo $\chi$
as indicated below.  To
spell out the details we extract material from \cite{gt,wa}
as follows.  We say more than is needed to display some of the beautiful
features of this subject and various connections to KdV are indicated
for possible further application to induced curves etc.
\\[3mm]\indent
Thus we go to \cite{cu,gt,wa} and note that $f\sim\bar{f}$ in transferring
results to the present context.
Let $(z,\bar{z})$ be coordinates on a Riemann
surface (RS).  A quasiconformal automorphism $z \to f(z,\bar{z}),
\bar{z} \to \bar{f}(z,\bar{z})$ is characterized by Beltrami coefficients
$\mu^z_{\bar{z}}, \mu_z^{\bar{z}}$ ($\mu = \mu^z_{\bar{z}})$ defined by
\be
\bar{\partial}f - \mu\partial f = 0\,\,(\mu =
\bar{\partial}f/\partial f)
\label{HA}
\ee
The Schwartz derivative (sometimes Schwartzian) is now
defined via
\be
S(f,z) = \{f,z\} = \frac{\partial^3 f}{\partial f}
- \frac{3}{2}(\frac{\partial^2 f}{\partial f})^2 = (log(\partial f)''
-\frac{1}{2}(log(\partial f)'^2
\label{HB}
\ee
and one knows that
\be
[\bar{\partial} - \mu\partial - 2(\partial \mu)]S(f,z) =
\partial^3 \mu
\label{HC}
\ee
(direct calculation - cf. \cite{cu} for a full evaluation).
This is an important result and comes up also in
connection with Ur-KdV following \cite{sb} - cf. also \cite{pe}).
Equation (\ref{HC}) is also present, but disguised, in \cite{cb,da} for
$C = \mu$ and $u\sim T = -\frac{1}{2}\kappa S_f$; the approach here
is much more meaningful and revealing.  Note here the minus sign in the
definition of $T$; this applies in this section only, and when comparing
with the $T$ of Section 3 we will keep matters clear.
Equation (\ref{HC}) can be rewritten
\be
\bar{\partial} T = [-\frac{1}{2}\kappa\partial^3
+ 2T\partial + T']\mu;\,\, T = -\frac{1}{2}\kappa S(f,z) = \kappa(p^2
+ p');\,\, p = -\frac{1}{2}\partial log(\partial f)
\label{HD}
\ee
(the distinction between $\partial g = g'$ or $g' + g\partial$
should be clear from context).
\\[3mm]\indent  We note a few calculations which are
needed below.  For $z \to f(z) = z + \epsilon(z)$ one has to first order
in $\epsilon, \epsilon',...,\,\,\,S = [(1+\epsilon')\epsilon''' -\frac{3}{2}
\epsilon''^2]/(1 + \epsilon')^2 \sim \epsilon'''$.  The transformation
law for T is $T \to f'^2T(f) + (c/12)S$ in conformal field theory (CFT)
so $\Delta T \sim (1+ \epsilon')^2(T + \epsilon T') - T + (c/12)\epsilon''' =
2\epsilon' T + \epsilon T' + (c/12)\epsilon'''$ (adjust c and $\kappa$
to bring notations into correspondence).  In (\ref{HI}) below in general one
needs infinitesimal maps $z \to z + \epsilon(z,\bar{z}),
\bar{z} \to \bar{z} + \bar{\epsilon}(z,\bar{z})$ and an expanded transformation
law. For such f, $\Delta f \sim f'\Delta z + \dot{f}\Delta \bar{z} \sim
f'\epsilon + \mu f' \bar{\epsilon}$ with $f' = 1$ at z ($\dot{f} \sim
\partial f/\partial \bar{z}$).  Thus, $\Delta f \sim \eta = \epsilon
+ \mu \bar{\epsilon}$ so $T \to (1+\eta')^2(T + \eta T')
-T + (c/12)\eta'''$ is consistent, which will imply (\ref{HI}) for $\delta T
\sim \Delta T$.  As for  generators we recall
(cf. \cite{ca}) that $T_{\epsilon} = \oint \epsilon(z)T(z)dz$ is called a
generator of conformal transformation where $[T_{\epsilon},T] \sim
\epsilon T' + 2\epsilon' T + (c/12)\epsilon'''$.  In our language one
can think of
\be
T_{\eta} = \int T(\zeta,\bar{\zeta})\eta(\zeta,\bar
{\zeta})d\zeta d\bar{\zeta};\,\, \{T(z,\bar{z}),T_{\eta}\} =
\label{HE}
\ee
$$\int \{T(z,\bar{z}),T(\zeta,\bar{\zeta})\}\eta(\zeta,\bar{\zeta})
d\zeta d\bar{\zeta} = \int\int\delta (\bar{z} - \bar{\zeta})
\hat{{\cal E}}_{z\bar{z}}\delta (z-\zeta)\eta d\zeta d\bar{\zeta} =
\hat{{\cal E}}\eta = \delta T$$
\noindent as a way of using a generator concept.  We refer here also to
\cite{ba,cb,da} where some of this information also
appears.  One can view
$\bar{z} \sim t$ and $z \sim x$ for example so (\ref{HD}) represents an
evolution equation for T (note $f,\bar{f} \sim$ independent variables,
$\hat{{\cal E}}\sim\tilde{{\cal E}} \sim -\lambda \partial^2 + T$, and u of KdV
$\sim$ EM tensor as indicated e.g. in \cite{ba,cb,da}).
\\[3mm]\indent Now to display this $\grave a$ la \cite{gt,wa}
we assume $\mu =
\delta H/\delta T$ for some Hamiltonian H and define the Poisson
brackets via
\be
\{P(z,\bar{z}),Q(z',\bar{z}')\} = \int
d \zeta d\bar{\zeta} (\delta P(z,\bar{z})/\delta T(\zeta,\bar{\zeta}))
\hat{{\cal E}}(\zeta,\bar{\zeta})(\delta Q(z',\bar{z}')/
\delta T(\zeta,\bar{\zeta}))
\label{HF}
\ee
where $\hat{{\cal E}} = -\frac{1}{2}\kappa \partial^3 + 2T\partial
+ T'$ (note that $T = T(z,\bar{z})$ in general now, not just $T = T(z)$).
For P = T and Q = H with $\mu = \delta H/\delta T$ one gets
\be
\{T,H\} = \{P,Q\} = \hat{{\cal E}}\mu\,\,(= \bar
{\partial} T)
\label{HG}
\ee
which expresses (\ref{HE}) in Hamiltonian form for $T=-\frac{1}{2}
\kappa S_f$ (note $\delta T(z,\bar{z})/\delta
T(\zeta,\bar{\zeta}) = \delta (z - \zeta)\delta (\bar{z} - \bar{\zeta})$).
For P = Q = T one has (note $\hat{{\cal E}}$ has no $\bar{\partial}$)
\be
\{T(z,\bar{z}),T(z',\bar{z}')\} = \delta (\bar{z} -
\bar{z}')\hat{{\cal E}}\delta (z - z') =
\delta (\bar{z} - \bar{z}')(-\frac{1}{2}\kappa\partial^3 + 2T\partial
+ T')\delta (z - z')
\label{HI}
\ee
To interpert this consider a quasiconformal diffeomorphism on
the RS: $z \to g(z,\bar{z}), \bar{z} \to \bar{g}(z,\bar{z})$ (these form
a quasiconformal group ${\cal G}$).  The infinitesimal form is
$z \to z + \epsilon(z,\bar{z}), \bar{z} \to \bar{z} + \bar{\epsilon}
(z,\bar{z})$ and T changes as ($\eta = \epsilon + \mu\bar{\epsilon})$
\be
\delta T = [-\frac{1}{2}\kappa \partial^3 +
2T\partial + T']\eta(z,\bar{z})
\label{HJ}
\ee
Then (\ref{HI}) means $T(z,\bar{z})$ is the
generator of $\eta$.  Finally
\be
\bar{\partial} f = \{f,H\} = \mu \partial f
\label{HK}
\ee
so the Beltrami equation is an evolution equation of f.  Here
one has
\be
\{f,H\} = \int (\delta f/\delta T)\hat{{\cal E}}
\mu d\zeta d\bar{\zeta} =\int (\delta f/\delta T)\bar{\partial} T
d\zeta d\bar{\zeta} \Rightarrow \{f,H\} = \bar{\partial} f
\label{HL}
\ee
since, via the generating function ideas of (\ref{HE}) ($\delta f =
\int\{f,T\}\eta d\zeta d\bar{\zeta}$) and the definition (\ref{HF}),
we have for $\eta = \mu\bar{\epsilon}$ and $\bar{\epsilon}\sim
\Delta\bar{z}$
\be
\delta f\sim\int \{f(z,\bar{z}),T(\zeta,\bar{\zeta})\}\eta(\zeta,
\bar{\zeta})d\zeta d\bar{\zeta}
\label{YY}
\ee
$$\sim\int\int\frac{\delta f(z,\bar{z})}{\delta T(\xi,\bar{\xi})}
\delta(\bar{\xi}-\bar{\zeta})\hat{{\cal E}}_{\xi}\delta(\xi - \zeta)
\eta d\xi d\bar{\xi} d\zeta d\bar{\zeta}\sim$$
$$\int\frac{\delta f(z,\bar{z})}{\delta T(\xi,\bar{\xi})}(\hat
{{\cal E}}\mu)\Delta\bar{z}d\xi d\bar{\xi}\sim\int
\frac{\delta f}{\delta T}\bar{\partial}T\Delta\bar{z}d\xi d\bar{\xi}$$
which leads to $\bar{\partial}f = \{f,H\}$.
\\[3mm]\indent Now for physics, in the light cone gauge, where $\bar{f}$
plays no role and the EM tensor $\sim$ Schwartzian as above,
we can find H such that $\mu = (\delta H/\delta T)$, so
(\ref{HG}),
i.e. $\bar{\partial} T = \{T,H\} = \hat{{\cal E}}\mu$, holds.
Under an
infinitesimal transformation $z \to z + \epsilon(z,\bar{z}), \bar{z}
\to \bar{z} + \bar{\epsilon}(z,\bar{z})$ we get formally (here
$d^2 x \sim (i/2)d\zeta d\bar{\zeta})$
\be
\delta H = \int d^2 x(\delta H/\delta T) \delta T =
\int d^2 x \mu\hat{{\cal E}}\eta = \int d^2 x \mu\hat{{\cal E}}
(\delta f/\partial f)
\label{HM}
\ee
($\delta f = \eta\partial f$ here).  Now integrate (\ref{HM}) to
get ($\dot{f} \sim \bar{\partial} f$)
\be
H = -\frac{\kappa}{4}\int d^2 x\frac{\dot{f}}
{f'}(\frac{f'''}{f'} - 2\frac{f''^2}{f'^2})
\label{HN}
\ee
which is a multiple of the geometric action of \cite{ag,ap}.
This integration is not clear in \cite{gt,wa} so we will sketch an
heuristic derivation as follows.  First from (\ref{HM})
\be
\delta H = \int d^2 x\mu\hat{{\cal E}}\eta =
\int d^2 x\eta\hat{{\cal E}}^{*}\mu = \int d^2 x(\frac{\delta f}{f'})
[\frac{1}{2}\kappa\partial^3 - 2\partial (T\,\,\cdot) + T']\mu
\label{HO}
\ee
$$= \int d^2 x(\frac{\delta f}{f'})[\frac{1}{2}\kappa\mu''' -2T'\mu
-2T\mu' + T'\mu] = -\int d^2 x(\frac{\delta f}{f'})\bar{\partial} T =
-\int d^2 x(\frac{\dot{T}}{f'})\delta f$$
Hence $(\delta H/\delta f) = -\frac{\dot{T}}{f'}$ (see (\ref{HT}) and
cf. \cite{cu} for computation of
$\dot{T}$), and we want an integral for H as in (\ref{HN}) where
\be
\delta H/\delta f = \sum (-\partial)^n(\delta H/\delta
f^{(n)}) - \bar{\partial} (\delta H/\delta \dot{f})
\label{HP}
\ee
in (\ref{HN}), where H is identified with the integrand also, in
a common abuse of notation.  To achieve this we write H in a slightly
different form in noting that
\be
\partial log(F')\bar{\partial} log(F') =
\frac{F''}{F'}\frac{\dot{F}'}{F'} \rightarrow -\dot{F}\partial
\frac{F''}{F'^2} = -\frac{\dot{F}}{F'}(\frac{F'''}{F'} -2\frac
{F''^2}{F'^2})
\label{HQ}
\ee
Here the arrow $\rightarrow$ indicates an integration by parts
where one assumes F and its derivatives are periodic or vanish suitably
at boundaries, or that the region has no boundary (i.e. the region is a
compact surface in which case the integrands must represent globally
defined objects).  In this situation
we can rewrite (\ref{HN}) as
\be
H = \frac{\kappa}{4}\int d^2 x\partial log(f')
\bar{\partial} log(f') = \frac{\kappa}{4}\int d^2 x(\frac{f''\dot{f}'}
{f'^2})
\label{HR}
\ee
\noindent Then using (\ref{HP}) (modulo $\kappa/4$)
\be
\delta H/\delta f \sim -\partial(-\frac{2f''\dot{f}'}
{f'^3}) + \partial \bar{\partial}(\frac{f''}{f'^2}) + \partial^2(\frac
{\dot{f}'}{f'^2})
= 2\frac{\dot{f}'''}{f'^2} - 6\frac{\dot{f}'' f''}{f'^3}
-2\frac{\dot{f}' f'''}{f'^3} + 6\frac{\dot{f}' f''^2}{f'^4}
\label{HS}
\ee
This is to compare with $-(\frac{\dot{T}}{f'})$ (modulo
$\kappa/4$) where
\be
-\frac{\dot{T}}{f'} =
-(\frac{\kappa}{4})[(\frac{-2}{f'})(\frac{\dot{f}'''}{f'} -\frac
{\dot{f}'f'''}{f'^2} -\frac{3}{2}(\frac{2\dot{f}'' f''}{f'^2}
-\frac{2\dot{f}' f''^2}{f'^3}))]
\label{HT}
\ee
Thus one has agreement without further integration by parts.
This procedure also gives a natural origin for geometric action, in
addition to the Virasoro algebra background of \cite{ag,aq,ap,cb}.
Let us summarize all this in (cf. \cite{gt,gu,gw,wa})
\\[3mm]\indent {\bf THEOREM 4.1}  First one has the equation $\bar{\partial} T
=\hat{{\cal E}}\mu$ of (\ref{HC}) - (\ref{HD}).  If there exists H such that
$\mu = \delta H/\delta T$,
one arrives at equations such as (\ref{HG}) $\bar{\partial} T = \{T,H\}$,
(\ref{HI}) $\{T(z,\bar{z}),T(z',\bar{z}')\} = \delta (\bar{z} - \bar{z}')
\hat{{\cal E}} \delta (z - z')$, and (\ref{HK}) $\bar{\partial} f =
\{f,H\} = \mu \partial f$. For transformations independent of $\bar{f}$
one can find $\mu =  (\delta H/\delta T)$ as in (\ref{HM}), (\ref{HO}) -
(\ref{HT}), leading to geometrical action as in (\ref{HN}).
\\[3mm]\indent {\bf REMARK 4.2}$\,\,$
Polyakov light cone action arises by
a simple calculation from H as in (\ref{HN}) for example via
(integration by parts - cf. \cite{wa} - we assume all terms are globally
defined)
\be
L = \int\mu T d\zeta d\bar{\zeta}  - H =
(\frac{\kappa}{4})\int d\zeta d\bar{\zeta}(\frac{\dot{f}'f''}{f'^2} -
\frac{\dot{f} f''^2}{f'^3})
\label{HU}
\ee
\indent {\bf REMARK 4.3}$\,\,$ One can arrive at a number of
connections of Liouville-Beltrami theory to KdV following
\cite{cu,gt,gu,gw,sg,sh,si,sl,wa}.  For example from (\ref{HC}) and (\ref{HG})
for $\mu = S(f,z)$ one obtains $\bar{\partial}\mu = \mu''' + 3\mu\mu'$
with a background Hamiltonian structure.  One can also consider a more
general Liouville-Beltrami action
\be
S = \int dz\wedge d\bar{z}(\frac{\dot{f}'f''}{f'^2}
-\frac{\dot{f}f''^2}{f'^3}) + \int dz\wedge d\bar{z}[\partial\phi
(\bar{\partial} - \mu\partial)\phi + \Lambda(e^{\phi}-1) + 2\phi
\partial^2 \mu]
\label{HH}
\ee
The first term is Polyakov's light cone action for 2-D gravity
 and the second term is Liouville action with a perturbation $2\phi
\partial^2\mu - \mu(\partial\phi)^2$, which is necessary for covariance
of the Beltrami-Liouville system.
Then such an action coupled with various matter fields leads to KdV
equations and bihamiltonian structure.
\\[3mm]\indent
We return now to the discussion at the beginning of Section 4
and look at $\Xi = S_{-}
+ S_{int}$ in (\ref{BJ}) written as ($f\to\bar{f},\,\,\mu =
\bar{f}_{\bar{z}}/\bar{f}_z$, etc. - we drop a factor $(i/2)$ arising
in the $z,\,\bar{z}$ integration)
\be
\Xi = \frac{k}{4\pi}\{\int\frac{\mu}{2}[\frac{\partial_z^3\bar{f}}
{\partial_z\bar{f}} - 2(\frac{\partial_z^2\bar{f}}{\partial_z\bar{f}})^2]
dz\wedge d\bar{z}
-\int\mu[\frac{\partial_z^3\bar{f}}{\partial_z\bar{f}} -\frac{3}{2}
(\frac{\partial_z^2\bar{f}}{\partial_z\bar{f}})^2]dz\wedge d\bar{z}\}
\label{IC}
\ee
(recall $T_{zz}\sim +S(\bar{f},z)$ in (\ref{BJ})).  But we see from
(\ref{HU}) that for $\kappa = k$
\be
\frac{1}{2\pi}L =
\frac{1}{2\pi} \int\mu T dz\wedge d\bar{z} -
 H \sim \frac{k}{4\pi}\int\mu
[\frac{1}{2}(\frac{\bar{f}'''}{\bar{f}'} - 2(\frac{\bar{f}''}{\bar{f}'})^2)-
\label{ID}
\ee
$$- (\frac{\bar{f}'''}{\bar{f}'} - \frac{3}{2}(\frac{\bar{f}''}{\bar{f}'})^2]
dz\wedge d\bar{z} = \Xi$$
(where $T\sim -(1/2)\kappa S(\bar{f},z)$ in (\ref{HU})).
Further from (\ref{HU}) one has then
\be
\frac{1}{2\pi}L =
\frac{1}{8\pi}\int\frac{\bar{f}_{zz}}{\bar{f}_z}(\frac
{\bar{f}_{z\bar{z}}}{\bar{f}_z} - \frac{\bar{f}_{zz}\bar{f}_{\bar{z}}}
{\bar{f}_z^2})dz\wedge d\bar{z} = -S_{+}
\label{IE}
\ee
which reduces $\Gamma_{eff} = S_{-} + S_{int} + S_{+}$
to zero (making a suitable extremum).  Also
$S_{+}\sim$ intrinsic Polyakov action so we have a direct calculation
showing
\\[3mm]\indent {\bf THEOREM 4.4}.$\,\,$ The equation of motion
(\ref{BK}) is satisfied for $f_1 = f_2 = \bar{f}$ and for this solution
$\Gamma_{eff} = 0$.
\\[3mm]\indent
Now it has been asserted previously that $S_{int}\sim\tilde{S}_P$ and
this means of course that $S_{int}\sim\tilde{S}_P$ modulo $\chi$ (since
from Theorem 3.4 for $h\sqrt{g} = 1$
one has $T_{zz} = 2p^2\bar{\mu}$ and $\mu T_{zz} =
2p^2|\mu|^2\not= p^2$ - cf. also (\ref{BA})).
To make this precise recall from (\ref{DH}) and (\ref{DJ}) that
(dropping $(i/2)$ again)
\be
\pi\chi\sim \int p^2(1 - |\mu|^2)dz\wedge d\bar{z};\,\,\tilde{S}_P \sim
\frac{4}{g_0^2}\int p^2dz\wedge d\bar{z}
\label{IF}
\ee
Hence in particular, writing $\Upsilon = \frac{g_0^2}{4}\tilde{S}_P
-\pi\chi$ we get
\be
\Upsilon = \int p^2|\mu|^2 dz\wedge d\bar{z} = \frac{1}{2}\int\mu T=
-\frac{2\pi}{k}S_{int}
\label{IG}
\ee
\indent {\bf THEOREM 4.5}.$\,\,$ For $h\sqrt{g} = 1$
the interaction term $\int\mu T_{zz} = -(4\pi/k)S_{int}$
above is in fact
$\int\mu T = 2\Upsilon = (g_0^2/2)\tilde{S}_P - 2\pi\chi$
and thus is equivalent to
the extrinsic Polyakov action modulo $\chi$
as stated previously without proof.  Note that $\Gamma_{eff} = 0\sim
\chi\not= 0$ generally and one could regard $\chi$ as fixed in determining
solutions.
\\[3mm]\indent {\bf REMARK 4.6}.$\,\,$  In \cite{pf} one wants a
dimensionless term in the action for string theory and this comes
from the extrinsic curvature which is added to $S_{NG}$.  This term
is unique (up to divergence terms) and invariant under scale transformations
$x\to\lambda x$.  It is essential to include this in the action and
one is led to a Grassman sigma model with constraint as in (\ref{AS})
or (\ref{BD}).  Such a string theory apparently corresponds to QCD.
In \cite{pc} one demonstrates that in ${\bf R}^3$ the
geometry of $h\sqrt{g} = 1$ surfaces is equivalent to WSO(3) gravity
but we have not developed this here.
\\[3mm]\indent We make also the following observation.
Take an extremal $\Gamma_{eff} = 0$ (cf. Theorem 4.3)
corresponding
to $0 = S_{-} + S_{+} + S_{int}$ with the same $\bar{f}$ ((\ref{BB}) is
satisfied).  Then $\int\mu T\sim
S_{-} + S_{+}$ (cf. and $\int\mu T\sim\tilde{S}_P$ up to a
factor of $\chi$ as in Theorem 4.5. We will argue then that
preservation of $\tilde{S}_P$ under mVN flows corresponds to preserving
the extremal WZNW action $\Gamma_{eff}$.  Now let us make this more
precise.  We have ($\Gamma$ and $S$ will be used interchangeably)
$\Gamma_{eff} = S_{+} + S_{-} + S_{int}$ with $\Gamma_{int} =
-(k/4\pi)\int\mu T$ (cf. (\ref{GY})).  Then via Theorem 4.5, $\int\mu T
= 2\Upsilon = (g_0^2/2)\tilde{S}_P - 2\pi\chi$.  On the other
hand, going to (\ref{ID}) we have $(1/2\pi)L = S_{-} - (k/4\pi)\int\mu T
= -S_{+}$.
Hence for common $\bar{f}$ one obtains
$S_{+} + S_{-} -(k/4\pi)\int\mu T = 0 =$
extreme $\Gamma_{eff}$  (recall
$\Gamma_{eff} = 0\sim\chi$ fixed).
Now under mVN flows with $h\sqrt{g} = 1,\,\,\tilde{S}_P$ is preserved, and
since the genus $g$ is integer valued, $\chi$
will be invariant under
continuous deformation. Consequently $\int\mu T$ and $S_{+} + S_{-}$
will also be preserved via the integration (\ref{HU}).  Hence
\\[3mm]\indent {\bf THEOREM 4.7}$\,\,$  Since $\tilde{S}_P$ is preserved
under mVN deformations with $h\sqrt{g} = 1$
via Theorem 5.1, let $\chi$ (preserved) be given;
then the extremal $\Gamma_
{eff} = 0$ equation (for $h\sqrt{g} = 1$) is also preserved, yielding
a family of extremal surfaces.

\section{MISCELLANEOUS RESULTS}
\renewcommand{\theequation}{5.\arabic{equation}}\setcounter{equation}{0}

We gather here various additional facts and observations which will be
organized as remarks.
\\[3mm]\indent {\bf REMARK 5.1}.$\,\,$
Now we recall also that $H_{zz} = \bar{\mu} = f_z/f_{\bar{z}}$ and
$H_{\bar{z}\bar{z}} = \mu = \bar{f}_{\bar{z}}/\bar{f}_z$ with
$T_{zz} = 2p^2\bar{\mu}$ and $\mu$ is given in (\ref{FG}).  The
formula (\ref{DG}) for $T_{zz}$ is more useful however since it implies
\be
T_{zz} = \frac{2}{\bar{\psi}_2}\partial_z(p\bar{\psi}_1) =
-\frac{2\bar{\psi}_{2zz}}{\bar{\psi}_2}
\label{GL}
\ee
This says that
\be
\bar{\psi}_{2zz} + \frac{T_{zz}}{2}\bar{\psi}_2 = 0
\label{GM}
\ee
which is a Schroedinger equation associated with KdV, with
$T_{zz}/2$ playing the role of potential.  It is tempting here to think
of $(1/2)T_{zz}\sim 2(log\tau)_{zz}$ by analogy to KdV but $T_{zz}$ here
is only defined to be a component of an EM tensor.  We note
from \cite{kk} that for geodesic
coordinates with $ds^2 = d\xi_1^2 + {\cal H}d\xi^2_2,\,\,{\cal H}^2
= G$ (i.e. $\sqrt{g}\sim {\cal H}$) with $S_{NG} = \sigma\int
\sqrt{g}d^2\xi$, one has
\be
{\cal H}_{\xi_1\xi_1} + K(\xi_1,\xi_2){\cal H} = 0
\label{GO}
\ee
where $K\sim$ Gaussian curvature.  Then ${\cal H}$ can be regarded
as $\Re\Psi$ where $\Psi$ is a wave function satisfying ($x\sim
\xi_1,\,\,y\sim\xi_2):\,\,-\Psi_{xx} + U(x,y)\Psi = \lambda^2\Psi$
for $K = -U +\lambda^2$.  There is a wide class of such $\Psi$ so many
surfaces could arise related to KdV in this manner.
Note that our $H_{\bar{z}\bar{z}}$ and other terms
involve conformal gauge but perhaps there is analogous behavior.  Note
first that in our notation $\sqrt{g} = 1/2p^2$ with $h = 2p^2$ for
$h\sqrt{g} = 1$.  Thus ${\cal H}\sim 1/2p^2$ in some sense, while
$H_{\bar{z}\bar{z}} = \mu$ plays the role of induced metric.  Here we
recall also from Theorem 3.4 that $\partial\mu = -\bar{\partial}\xi
\equiv \mu_z = 2(log(p))_{\bar{z}}$ and $\bar{\partial}\bar{\mu} = -\partial\xi
\equiv \bar{\mu}_{\bar{z}} = 2(log(p))_z$ while the Liouville equation states
that $\xi_{z\bar{z}} = -Kexp(\xi)\sim (log(p))_{z\bar{z}} = K/4p^2$.
\\[3mm]\indent {\bf REMARK 5.2}.$\,\,$
There is also another interesting
direction suggested by some formulas in \cite{ho}.  Thus given a map
$g:\,M\mapsto N;\,\,w = f(z)$, between two surfaces with conformal metrics
$ds^2 = \lambda^2|dz|^2$ and $d\sigma^2 = \nu^2|dw|^2$ respectively,
one defines an energy density $e(g) = (1/2)\|dg\|^2 = e'(g) + e''(g)$
where $e'\sim(\nu^2/\lambda^2)|w_z|^2$ and $e''\sim(\nu^2/\lambda^2)
|w_{\bar{z}}|^2$.  For $D\subset M$ the total energy is $E(g) =
\int_D e(g)d^2\xi$ and the associated Euler-Lagrange operator is called
the tension field
\be
\tau(g) = \frac{4}{\lambda^2}(w_{z\bar{z}} + 2(log\nu)_w w_z w_{\bar{z}})
\label{JV}
\ee
For $N = S^2(R) =$ sphere of radius $R$ and $w$ a conformal parameter
obtained by a similarity transformation $S^2(R)\mapsto S^2(1)$ followed
by a stereographic projection one has $\nu = 2R/(1 + |w|^2)$.  Then if
$M$ is a plane domain ($\lambda = 1$) one obtains ($w\sim f(z,\bar{z})$)
\be
e'(g) = 4R^2\frac{|f_z|^2}{(1 + |f|^2)^2};\,\,e''(g) = 4R^2
\frac{|f_{\bar{z}}|^2}{(1 + |f|^2)^2}
\label{JW}
\ee
$$\tau(g) = 4(f_{z\bar{z}} - \frac{2\bar{f}f_z f_{\bar{z}}}{1 + |f|^2}) =
4\ell(f)$$
Setting $F(f) = f_{\bar{z}}/(1 + |f|^2)$ and $\hat{F}(f) = f_z/
(1+ |f|^2)$ we see that $g$ holomorphic $\sim F\equiv 0$, $g$ antiholomorphic
$\sim \hat{F}\equiv 0$, and $g$ harmonic $\sim\ell(f)\equiv 0$.  This all
leads to some interesting relations among the various actions studied
earlier.  Thus write (h and ah for holomorphic and antiholomorphic
respectively)
\be
S_{ah} = \int e''(g)d^2\xi;\,\,S_h = \int e'(g)d^2\xi
\label{JX}
\ee
Now via (\ref{DH}) we see that (set $d^2\xi = -(2/i)dz\wedge d\bar{z}$)
\be
S_{ah} = -4R^2\frac{2}{i}\int\frac{|f_{\bar{z}}|^2}{(1+|f|^2)^2}dz\wedge
d\bar{z} = 4R^2g_0^2\tilde{S}_P
\label{JY}
\ee
and from (\ref{AU})
\be
2\pi\chi = \frac{i}{4R^2}\int (e'' - e')dz\wedge d\bar{z} =
\frac{1}{8R^2}(S_{ah} - S_h)
\label{JZ}
\ee
We note that the notation in \cite{ho} corresponds to an interior normal
vector so a minus sign adjustment may be needed at times.  Further we
observe that $df\wedge d\bar{f} = (|f_z|^2 - |f_{\bar{z}}|^2)dz\wedge
d\bar{z}$ which leads to
\be
2\pi\chi = i\int\frac{df\wedge d\bar{f}}{(1 + |f|^2)^2}
\label{KA}
\ee
\\[3mm]\indent {\bf THEOREM 5.3}.$\,\,$ With no restriction
$h\sqrt{g} = 1$ we get (\ref{JY}) - (\ref{KA}).
\\[3mm]\indent {\bf REMARK 5.4}$\,\,$
One has a direct connection of the context of Remark 5.2 to the classical
2-D $SO(3)$ sigma model following \cite{pz,ra}.  Here (with appropriate
variables and scaling) $S\sim S_h + S_{ah}$ corresponds to the sigma
model action and the charge $Q\sim S_h - S_{ah}\sim -\chi$.  The equation
of motion corresponds then to (using $f\sim w$ in \cite{pz} - $\bar{f}\sim
w$ could also be used)
\be
f_{z\bar{z}} = \frac{2\bar{f}f_z f_{\bar{z}}}{1 +|f|^2}
\label{PA}
\ee
(cf. (\ref{AS}) and (\ref{BD}) and note this corresponds to $h_z
f_{\bar{z}}/h = 0$ in (\ref{QQ})).
One finds that $S\geq 4\pi|Q|$ and it turns out that multiinstanton
(or antiinstanton) solutions of the form
\be
f = \prod_1^k\frac{z-a_j}{z-b_j};\,\,f = \prod_1^k\frac{\bar{z} - a_j}
{\bar{z} - b_j}
\label{PB}
\ee
of charge $\pm k$
are the only solutions of (\ref{PA}) with finite action.
\\[3mm]\indent {\bf REMARK 5.5}.$\,\,$
The formulas (\ref{DH}), (\ref{DI}) for actions as well as (\ref{DG})
and (\ref{DJ}) for $T_{zz}$ and $\chi$
respectively have an interesting structure
and one can in fact develop this further.  First let us think
of $\psi_1$ and $\psi_2$ with their conjugates as determining a map
$m:\,(z,\bar{z})\mapsto (\psi_1,\psi_2):\,
{\bf C}\to {\bf C}^2$ (one could also envision $(z,\bar{z})\mapsto
(\psi_1,\bar{\psi}_1,\psi_2,\bar{\psi}_2)$ but this is somewhat less
clear).  Via the defining equations (*) $\psi_{1z} =
p\psi_2,\,\,\psi_{2\bar{z}} = -p\psi_1$ (and $\bar{\psi}_{1\bar{z}}
= p\bar{\psi}_2,\,\bar{\psi}_{2z} = -p\bar{\psi}_1$) one has constraints
\be
(\psi_1^2)_z + (\psi_2^2)_{\bar{z}} = 0;\,\,\frac{\bar{\psi}_{1\bar{z}}}
{\bar{\psi}_2} = \frac{\psi_{1z}}{\psi_2};\,\,\frac{\bar{\psi}_{2z}}
{\bar{\psi}_1} = \frac{\psi_{2\bar{z}}}{\psi_1}
\label{GA}
\ee
Observe that $\alpha(\psi_1,\psi_2)$ satisfies
the same constraints for $\alpha\in {\bf C}$,
so we can think of $m:\,{\bf C}\to  {\bf C}P^1
= {\bf P}C^1\simeq S^2$ (Riemann sphere) directly, without using
(\ref{AR}), (\ref{DA}), etc.  Now consider the extrinsic Polyakov action
$\tilde{S}_P$ of (\ref{DH}) in the form $\tilde{S}_P = (2i/g_0^2)\int
p^2 dz\wedge d\bar{z}$ ($h\sqrt{g} = 1$).
One can rewrite this in terms of the $\psi_i$
as follows.  Using $p = (\psi_{1z}/\psi_2) = (\bar{\psi}_{1\bar{z}}/\bar
{\psi}_2)$ one has
\be
\tilde{S}_P = \frac{2i}{g_0^2}\int\frac{\psi_{1z}\bar{\psi}_{1\bar{z}}}
{|\psi_2|^2}dz\wedge d\bar{z}
\label{GB}
\ee
Now use $d\psi_1 = \psi_{1z}dz + \psi_{1\bar{z}}d\bar{z}$ and
$d\bar{\psi}_1 = \bar{\psi}_{1z}dz + \bar{\psi}_{1\bar{z}}d\bar{z}$
to get $d\psi_1\wedge d\bar{\psi}_1 = (\psi_{1z}\bar{\psi}_{1\bar{z}}
-\psi_{1\bar{z}}\bar{\psi}_{1z})dz\wedge d\bar{z}$ and write
\be
\tilde{S}_P = \frac{2i}{g_0^2}\int \frac{d\psi_1\wedge d\bar{\psi}_1}
{|\psi_2|^2} + \frac{2i}{g_0^2}\int\frac{\psi_{1\bar{z}}\bar{\psi}_{1z}}
{|\psi_2|^2}dz\wedge d\bar{z}
\label{GC}
\ee
Our basic equations say nothing about $\psi_{1\bar{z}}$ or $\bar{\psi}_{1z}$
however and a change of variables $z\leftrightarrow\bar{z}$ in the
second integral does not give a copy of $\tilde{S}_P$ since e.g.
$\psi_1(z,\bar{z})_{\bar{z}}\to\psi_1(\bar{z},z)_z\not=\psi_1(z,\bar{z})$
without further hypotheses.  Another approach would be to write $p^2 =
-(\psi_{1z}/\psi_2)(\psi_{2\bar{z}}/\psi_1) = -(log\psi_1)_z(log\psi_2)_
{\bar{z}}$ and then
\be
\tilde{S}_P = -\frac{2i}{g_0^2}\int (log\psi_1)_z(log\psi_2)_{\bar{z}}
dz\wedge d\bar{z}
\label{GD}
\ee
Setting $d(log\psi_1) = (log\psi_1)_zdz + (log\psi_1)_{\bar{z}}d\bar{z}$ and
$d(log\psi_2) = (log\psi_2)_z dz + (log\psi_2)_{\bar{z}}d\bar{z}$ we get
$d(log\psi_1)\wedge d(log\psi_2) = [(log\psi_1)_z(log\psi_2)_{\bar{z}}
-(log\psi_1)_{\bar{z}}(log\psi_2)_z]dz\wedge d\bar{z}$ leading to
\be
\tilde{S}_P = -\frac{2i}{g_0^2}\int d(log\psi_1)\wedge d(log\psi_2)
-\frac{2i}{g_0^2}\int (log\psi_1)_{\bar{z}}(log\psi_2)_z dz\wedge d\bar{z}
\label{GE}
\ee
and again terms $\psi_{1\bar{z}}/\psi_1$ and $\psi_{2z}/\psi_2$ are not
specified a priori.
\\[3mm]\indent
Suppose now that we require, in addition to the basic equations (*)
$\psi_{1z} = p\psi_2,\,\,\psi_{2\bar{z}} = -p\psi_1$, that
\be
\psi_{1\bar{z}} = q\psi_2;\,\,\psi_{2z} = -q\psi_1
\label{GF}
\ee
for some function $q(z,\bar{z})$.  Then $\bar{\psi}_{1z} = \bar{q}\bar
{\psi}_2$ and $\bar{\psi}_{2\bar{z}} = -\bar{q}\bar{\psi}_1$ so that
all derivatives of the $\psi_i$ would be specified.  In particular
consider the last term in (\ref{GC}) with integrand $(\psi_{1\bar{z}}/
\psi_2)(\bar{\psi}_{1z}/\bar{\psi}_2) = |q|^2(z,\bar{z})$ and suppose
$|q|^2(z,\bar{z}) = p^2(\bar{z},z)$.  Then for $z\leftrightarrow\bar{z},\,\,
\int |q|^2(z,\bar{z})dz\wedge d\bar{z}\to\int p^2(z,\bar{z})d\bar{z}
\wedge dz = -\int p^2(z,\bar{z})dz\wedge d\bar{z}$ and (\ref{GC}) yields
\be
\tilde{S}_P = \frac{i}{g_0^2}\int\frac{d\psi_1\wedge d\bar{\psi}_1}{|\psi_2|^2}
\label{GG}
\ee
Similarly in (\ref{GF}) $(\psi_{1\bar{z}}/\psi_1)(\psi_{2z}/\psi_2) =
-q^2(z,\bar{z})$ so if we take $q$ real with $-q^2(z,\bar{z}) =
-p^2(\bar{z},z)$ then the last integral in (\ref{GE}) becomes
$-\int q^2(z,\bar{z})dz\wedge d\bar{z} = -\int p^2(\bar{z},z)dz\wedge
d\bar{z} = -\int p^2(z,\bar{z})d\bar{z}\wedge dz = \int p^2(z,\bar{z})
dz\wedge d\bar{z}$ so that (\ref{GE}) becomes
\be
\tilde{S}_P = -\frac{i}{g_0^2}\int d(log\psi_1)\wedge d(log\psi_2)
\label{GH}
\ee
If $q$ is real in (\ref{GF}) then both (\ref{GG}) and (\ref{GH}) hold.
We can also write from (*) $(\psi_1^2)_z + (\psi_2^2)_{\bar{z}} = 0$ while
from (\ref{GF}) one has $(\psi_1^2)_{\bar{z}} + (\psi_2^2)_z = 0$.
Adding and subtracting leads to
\be
(\partial_z + \partial_{\bar{z}})(\psi_1^2 + \psi_2^2) = 0;\,\,
(\partial_z - \partial_{\bar{z}})(\psi_1^2 - \psi_2^2) = 0
\label{TA}
\ee
Hence for ($\alpha,\beta$) arbitrary functions
\be
\psi_1^2 = \alpha(z+\bar{z}) + \beta(z-\bar{z});\,\,\psi_2^2 =
\alpha(z+\bar{z}) - \beta(z-\bar{z})
\label{TB}
\ee
We note a few additional formulas which hold when (\ref{GF}) applies.
Thus from (\ref{FG})
\be
\mu = H_{\bar{z}\bar{z}} = 2(\bar{\psi}_2\psi_{1\bar{z}} -
\psi_1\bar{\psi}_{2\bar{z}}) = 2q(|\psi_1|^2 + |\psi_2|^2) = \frac{q}{p}
\label{GR}
\ee
so $\mu$ is real, while from (\ref{FF}) and Theorem 3.2
\be
T_{zz} = 4p^2(\bar{\psi}_{1z}\psi_2 - \bar{\psi}_1\psi_{2z}) =
4p^2 q(|\psi_1|^2 + |\psi_2|^2) = 2pq
\label{GS}
\ee
If we write (\ref{GR}) and (\ref{GS}) in terms of the
$\psi_i$ one obtains
\be
T_{zz} = 2pq = -2\partial log(\psi_1)\partial log(\psi_2);\,\,
H_{\bar{z}\bar{z}} = \frac{q}{p} = -\frac{\bar{\partial}\psi_1^2}
{\bar{\partial}\psi_2^2}
\label{GU}
\ee
Thus we have proved
\\[3mm]\indent {\bf THEOREM 5.6}$\,\,$
Given (\ref{GF}) (plus (*)) one obtains (\ref{TB}) for arbitrary
($\alpha,\beta$) which can be written as $\psi_1^2 = \alpha(x) +\beta(y)$
and $\psi_2^2 = \alpha(x) -\beta(y)$.  Further (given (*)),
if (\ref{GF}) holds with
$q(z,\bar{z}) = p(\bar{z},z)$ real then (\ref{GG}) and (\ref{GH}) hold
for $\tilde{S}_P$.
Thus the lovely formulas (\ref{GG}) - (\ref{GH}) occur for the particular
class of surfaces for which (cf. (\ref{AI}))
\be
\partial_z X^1 = i(\alpha + \bar{\alpha} + \bar{\beta}-\beta);\,\,
\partial_z X^2 = \bar{\alpha} + \bar{\beta} -\alpha + \beta;
\label{TC}
\ee
$$\partial_z X^3 = -2\sqrt{(\alpha - \beta)(\bar{\alpha} + \bar{\beta})};\,\,
g_{12} = 4[|\alpha|^2 + |\beta|^2 + |\alpha + \beta||\alpha - \beta|]$$
Further one obtains (\ref{GR}) - (\ref{GU}).
\\[3mm]\indent {\bf REMARK 5.7}.$\,\,$
We note in passing also that when (\ref{GF}) holds one gets (cf.
(\ref{GG}))
\be
S_{ah} = 8iR^2\int p^2 dz\wedge d\bar{z};\,\, S_h = 8iR^2\int
q^2 dz\wedge d\bar{z} = -S_{ah}
\label{KB}
\ee
so that from (\ref{AU})
\be
2\pi\chi = \frac{1}{4R^2}S_{ah} = g^2_0\tilde{S}_P
\label{KC}
\ee
Also in connection with (\ref{GF}) we note that (\ref{GF})
implies $\mu$ is real from (\ref{GR}) while $T_{zz}$ is then automatically
real from (\ref{GS}).  Conversely if $\mu$ is real one must have
\be
\bar{\psi}_2\psi_{1\bar{z}} - \psi_1\bar{\psi}_{2\bar{z}} =
\psi_2\bar{\psi}_{1z} - \bar{\psi}_1\psi_{2z}
\label{KD}
\ee
This can happen in at least two ways, namely
\be
\frac{\psi_{1\bar{z}}}{\psi_2} = q;\,\,\frac{\psi_{2z}}{\psi_1} = r\,\
or\,\,\frac{\psi_{1\bar{z}}}{\bar{\psi}_1} = -\frac{\psi_{2z}}
{\bar{\psi}_2}
\label{KE}
\ee
with $(q,r)$ real ($r = -q\sim$ (\ref{GF})).
We see from $T_{zz} = 2p^2\bar{\mu}$ in Theorem 3.4 that $T_{zz}$ is
now automatically real with $\mu$.  Further from (\ref{FF}) and (\ref{KE})
one has
\be
\bar{\psi}_1\psi_{1\bar{z}} + \bar{\psi}_{2\bar{z}}\psi_2 = \frac{1}{2}
\partial(\frac{1}{p})\Rightarrow \bar{\psi}_1\psi_2 (q+ r) =
\frac{1}{2}\partial(\frac{1}{p})
\label{KF}
\ee
which implies $\bar{\psi}_1\psi_2$ is real.  On the other hand $\mu$ real
implies $\partial\mu = 2\bar{\partial}log(p)$ and $\bar{\partial}\mu =
2\partial log(p)$ which implies $\partial^2 log(p) = \bar{\partial}^2
log(p)$ and consequently
\be
log(p) = F(z+\bar{z}) + G(z-\bar{z})\sim p = f(z+\bar{z})g(z-\bar{z})
\label{KG}
\ee
Apparently however $r = -q$ in (\ref{KE}) is not implied by $\mu$ real.
Thus in particular the condition (\ref{GF}) seems to be rather strong.
\\[3mm]\indent {\bf REMARK 5.8}.$\,\,$
The quantities in $A'_z$ and $A'_{\bar{z}}$ of (\ref{AZ}) can be computed
for $h\sqrt{g} = 1$.  Thus in $A'_z$ one has entries $\pm(1/\sqrt{2})(1+T)$
and $\pm(i/\sqrt{2})(1-T)$ with $T = 2p^2\bar{\mu}\,\,(\mu = 2\bar{\psi}^2_2
(\psi_1/\bar{\psi}_2)_{\bar{z}})$,
while in $A'_{\bar{z}}$
we have $\pm(1/\sqrt{2})(\mu + 2p^2),\,\,\pm(1/\sqrt{2})(\mu - 2p^2),$
and $\pm i(log\,p^2)_{\bar{z}}$.  The components for the $\hat{e}_i$ in
(\ref{AY}) are
\be
\Lambda =\frac{1}{1+|f|^2} = 2p|\psi_2|^2;\,\,\Lambda(f^2 + \bar{f}^2) =
-4p\Re (\frac{\psi_1^2\psi_2}{\bar{\psi}_2});
\label{KH}
\ee
$$ \Lambda(f^2 - \bar{f}^2) = 4ip\Im (\frac{\psi_1^2\psi_2}{\bar{\psi}_2});\,\,
\Lambda  (f+\bar{f}) = 4p\Im (\psi_1\psi_2);\,\,\Lambda(f-\bar{f})=
4ip\Re (\psi_1\psi_2)$$
$$\Lambda[1+\frac{1}{2}(f^2 + \bar{f}^2)] = 2p[|\psi_2|^2
-\Re (\frac{\psi_1\psi_2}{\bar{\psi}_2})]$$

\end{document}